\documentclass[twocolumn,longauth]{aa}  

\usepackage{amsmath}
\usepackage{amssymb}

\usepackage{graphicx}
\usepackage{txfonts}
\usepackage{xcolor}

\usepackage{bm}
\renewcommand{\vec}[1]{\bm{#1}}

\usepackage[]{hyperref}
\newcommand{\base}[2]{{\ensuremath{#1, #2}}}

\newcommand{\pinv}{{\ensuremath{\mathrm{+}}}}

\newcommand{\real}[1]{{\ensuremath{\mathrm{Re}\left(#1\right)}}}

\newcommand{\mean}[1]{{\ensuremath{\left<#1\right>}}}

\begin{document}

\title{Upgrading the GRAVITY fringe tracker for GRAVITY+}
\subtitle{Tracking the white light fringe in the non-observable Optical Path Length state-space}

\author{
M. Nowak \inst{\ref{ioa}, \ref{kavli}}\thanks{These authors contributed equally.}
\and 
S. Lacour \inst{\ref{lesia}, \ref{eso}}$^\star$
\and 
R. Abuter \inst{\ref{eso}}
\and 
J. Woillez \inst{\ref{eso}}
\and 
R. Dembet \inst{\ref{lesia}}
\and
M. S.    Bordoni \inst{\ref{mpe}}
\and
G.    Bourdarot \inst{\ref{mpe}}
\and
B.  Courtney-Barrer \inst{\ref{eso}, \ref{anu}}
\and
D.    Defr\`ere \inst{\ref{leuven}}
\and
A.    Drescher \inst{\ref{mpe}}
\and
F.    Eisenhauer \inst{\ref{mpe}}
\and
M.    Fabricius \inst{\ref{mpe}}
\and
H.    Feuchtgruber \inst{\ref{mpe}}
\and
R. Frahm \inst{\ref{eso}}
\and
P.    Garcia \inst{\ref{porto}, \ref{lisboa}}
\and
S.    Gillessen \inst{\ref{mpe}}
\and
V.    Gopinath \inst{\ref{mpe}}
\and
J.     Graf \inst{\ref{mpe}}
\and
S.    Hoenig \inst{\ref{soton}}
\and
L.    Kreidberg \inst{\ref{mpia}}
\and
R.    Laugier \inst{\ref{leuven}}
\and
J.B.    Le Bouquin \inst{\ref{ipag}}
\and
D.    Lutz \inst{\ref{mpe}}
\and
F.    Mang \inst{\ref{mpe}}
\and
F.    Millour \inst{\ref{oca}}
\and
N.    More \inst{\ref{mpe}}
\and
N. Moruj\~{a}o \inst{\ref{porto}, \ref{lisboa}}
\and
T.    Ott \inst{\ref{mpe}}
\and
T.    Paumard \inst{\ref{lesia}}
\and
G.    Perrin \inst{\ref{lesia}}
\and
C.    Rau \inst{\ref{mpe}}
\and
D. C. Ribeiro \inst{\ref{mpe}}
\and
J.    Shangguan \inst{\ref{mpe}}
\and
T.    Shimizu \inst{\ref{mpe}}
\and
F.    Soulez \inst{\ref{cral}}
\and
C.    Straubmeier \inst{\ref{koeln}}
\and
F.    Widmann \inst{\ref{mpe}}
\and
B. Wolff \inst{\ref{eso}}
}

\institute{
  Institute of Astronomy, University of Cambridge, Madingley Road, Cambridge CB3 0HA, United Kingdom   \label{ioa}      \\
  \email{mcn35@cam.ac.uk}  
  \and  
  Kavli Institute for Cosmology, University of Cambridge, Madingley Road, Cambridge CB3 0HA, United-Kingdom \label{kavli}
  \and
  LESIA, Observatoire de Paris, Universit\'e PSL, CNRS, Sorbonne Universit\'e, Univ. Paris Diderot, Sorbonne Paris Cit\'e, 5 place Jules Janssen, 92195 Meudon, France \label{lesia}
  \and 
  European Southern Observatory, Karl-Schwarzschild-Straße 2, 85748 Garching, Germany \label{eso}
  \and 
  Max Planck Institute for extraterrestrial Physics, Giessenbachstraße 1, 85748 Garching, Germany \label{mpe}
  \and 
  Research School of Astronomy \& Astrophysics, Australian National University, ACT 2611, Australia \label{anu}
  \and 
  Department of Physics \& Astronomy, University of Southampton, Hampshire, SO17 1BJ, Southampton, UK \label{soton}
  \and 
  Max Planck Institute for Astronomy, Konigstuhl 17, 69117 Heidelberg, Germany \label{mpia}
  \and 
  Univ. Grenoble Alpes, CNRS, IPAG, 38000 Grenoble, France \label{ipag}
  \and 
  Institute of Astronomy, KU Leuven, Celestijnenlaan 200D, B-3001 Leuven,
Belgium \label{leuven}
  \and 
  Laboratoire Lagrange, Observatoire de la Cote d'Azur, CNRS, Boulevard de l'Observatoire, CS 34229, 06304 Nice, France \label{oca}
  \and 
  Universite de Lyon, CNRS, Centre de Recherche Astrophysique de Lyon UMR 5574, 69230 Saint-Genis-Laval, France \label{cral}
  \and 
  Universidade do Porto, Faculdade de Engenharia, Rua Dr. Roberto Frias, 4200-465 Porto, Portugal \label{porto}
  \and
CENTRA - Centro de Astrof\'isica e Gravita\c{c}\~{a}o, IST, Universidade de Lisboa, 1049-001 Lisboa, Portugal \label{lisboa}
  \and  
  1. Institute of Physics, University of Cologne, Z\"ulpicher Straße 77, 50937 Cologne, Germany \label{koeln}
}
             
\date{\today}
   
\titlerunning{Upgrading the GRAVITY fringe tracker}
   
\abstract
    {Over the recent years, the GRAVITY beam-combiner at the Very Large Telescope Interferometer has made important contributions to many different fields of astronomy, from observations of the Galactic centre to the study of massive stars, young stellar objects, exoplanet atmospheres, and active galactic nuclei. Getting to this point where optical interferometry is playing a pivotal role in astronomy was only made possible by the development of several key technologies. Among these, the development of a reliable and high-performance fringe tracker stands out. 
    These systems compensate for disturbances ranging from atmospheric turbulence to vibrations in the optical system, enabling long exposures and ensuring the stability of interferometric measurements.
    }
    {As part of the ongoing GRAVITY+ upgrade of the Very Large Telescope Interferometer infrastructure, we aim to improve the performance of the GRAVITY Fringe-Tracker, and to enable its use by other instruments.}
    {We modify the group delay controller to consistently maintain tracking in the white light fringe, characterised by a minimum group delay. Additionally, we introduce a novel approach in which fringe-tracking is performed in the non-observable Optical Path Length state-space, using a covariance-weighted Kalman filter and an auto-regressive model of the disturbance. We outline this new state-space representation, and the formalism we use to propagate the state-vector and generate the control signal. While our approach is presented specifically in the context of GRAVITY/GRAVITY+, it can easily be adapted to other instruments or interferometric facilities.}
    {We successfully demonstrate phase delay tracking within a single fringe, with any spurious phase jumps detected and corrected in less than 100\,ms. We also report a significant performance improvement, as evidenced by a reduction of $\sim{}30$ to $40\%$ in phase residuals, and a much better behaviour under sub-optimal atmospheric conditions. Compared to what was observed in 2019, the median residuals have decreased from 150~nm to 100~nm on the Auxiliary Telescopes and from 250~nm to 150~nm on the Unit Telescopes.}
    {The improved phase-delay tracking combined with whit light fringe tracking means that from now-on, the GRAVITY Fringe-Tracker can be used by other instruments operating in different wavebands. The only limitation remains the need for an optical path dispersion adjustment.}


\maketitle

\section{Introduction}

By combining the light from multiple telescopes, long-baseline optical interferometry can achieve angular-resolution much higher that what is possible with a single-dish telescope.
 Even in the upcoming era of 30+m class telescope, the Very Large Telescope Interferometer (VLTI) will continue to outperform single-dish observations in terms of angular resolution \citep[for an up-to-date overview, see][]{2023ARA&A..61..237E}. This is particularly useful for many applications, as evidence by the broad range of subjects to which it has been applied with groundbreaking results \citep{2018A&A...615L..15G,2018A&A...618L..10G,2018Natur.563..657G,2019A&A...625L..10G,2019A&A...623L..11G,2019A&A...632A..53G,2020A&A...636L...5G,2020A&A...643A.154G,2020A&A...635A.143G,2020A&A...635A..92G,2020A&A...633A.110G,2020A&A...634A...1G,2021A&A...647A..59G,2022A&A...657L..12G}.

One of the challenges which arise at such high angular resolution is the need to manage disturbances which can include atmospheric turbulence, telescope vibrations, and other noise sources, and which remain the primary limitations to the precision of interferometric measurements. Such disturbances lead to variations in the Optical Path Length (OPL) of light travelling through each telescope to the beam combiner, resulting in an unstable interference pattern (unstable "fringes").

To obtain stable interferometric measurements, these fringes need to be controlled in real-time, and any OPL fluctuations need to be compensated for. In modern instruments, this is the role of a specific subsystem: the Fringe-Tracker (FT). The crucial role of Fringe-Tracking in optical interferometry has justified the extensive development efforts made over the past decades \citep{Shao1988,Sorrente2001,2006SPIE.6268E..0UD,2008A&A...481..553L, Cassaing-p-08a, Houairi-p-08a,2010PASP..122..795C,Lozi-p-11, 2012A&A...541A..81M,  2014A&A...569A...2C,2019A&A...624A..99L}. Various techniques have been proposed, either based on hardware or software solutions, but they all share a common objective: to measure and stabilise the fringes, thereby maximising the Signal-to-Noise Ratio (SNR) of interferometric measurements.

For the GRAVITY instrument, the fringe-tracker was initially designed to perform control in the 6-dimensional Optical Path Delay (OPD) state-space, where measurements are available. The algorithm, described in \cite{2019A&A...624A..99L} (subsequently Paper I), was based around the use of an Auto-Regressive (AR) model to model the disturbance in OPD space, and a Kalman filter to propagate an OPD state-vector using an asymptotic approximation of the optimal gain.

Deviating from this initial concept, we hereafter introduce a novel approach in which fringe-tracking is performed in the 4-dimensional non-observable OPL state-space. The reduction in dimensionality alleviates the need for an asymptotic approximation of the Kalman gain, whose optimal value is now dynamically calculated at each iteration. This work is part of a general upgrade of the GRAVITY Fringe-Tracker which happened in late 2022 and 2023, and can be seen as a follow-up to Paper~1.

The paper is divided as follows. Section~\ref{sec:observables} gives a general description of how measurements are made within the FT, and which quantities are observable. Section~\ref{sec:control_loop} describes the details of the implementation of the upgraded FT, and in particular the state-space model, and the two control loops (phase and group-delay). Section~\ref{sec:onsky} reports on the first on-sky results obtained with this upgraded FT, with an emphasis on the behaviour of the group-delay control loop. Finally, Section~\ref{sec:end} gives some prospects and avenues for future work, and Section~\ref{sec:conclusion} gives our final conclusions.

\section{Observables}
\label{sec:observables}

\subsection{From pixels to fluxes}

In GRAVITY, the fringe tracking beam combiner, like the science combiner, is a silica-on-silicium integrated optics component, which uses an ABCD beam-combining scheme. This means that for any given baseline $\base{j}{k}$ linking telescopes $T_j$ to $T_k$, the beam-combiner registers a set of four outputs, corresponding to four values of additional phase shifts introduced between the beams (0, $\pi/2$, $\pi$, and $3\pi/2$):
\begin{align}
  a_\base{j}{k} &= \left<|E_j|^2 + |E_k|^2 + 2\real{E_jE_k^*}\right> ,\label{eq:a}\\
  b_\base{j}{k} &= \left<|E_j|^2 + |E_k|^2 + 2\real{E_jE_k^*e^{i\pi/2}}\right> ,\\
  c_\base{j}{k} &= \left<|E_j|^2 + |E_k|^2 + 2\real{E_jE_k^*e^{i\pi}}\right> ,\\
  d_\base{j}{k} &= \left<|E_j|^2 + |E_k|^2 + 2\real{E_jE_k^*e^{3\pi/2}}\right> \label{eq:d}
,\end{align}
\noindent{}where $E$ is the complex amplitude of the electric field at the input of the combiner, $\left<.\right>$ denotes the average over the integration time, and $^*$ the complex conjugate.

In principle, both the coherent flux $\Gamma_\base{j}{k}=\mean{E_jE_k^*}$ and the incoherent flux $F_j + F_k = \mean{|E_j|^2} + \mean{|E_k|^2}$ can be recovered by linear combinations of these 4 outputs:
\begin{align}
  F_j + F_k = \left(a_\base{j}{k} + b_\base{j}{k} + c_\base{j}{k} + d_\base{j}{k}\right)/4,\\
  \Gamma_\base{j}{k} = (a_\base{j}{k}-c_\base{j}{k})/2 + i\,(d_\base{j}{k} - b_\base{j}{k})/2
.\end{align}
\noindent{}And each individual $F_j$ for $j\in{}[1\dots{}4]$ can subsequently be recovered by a second linear combination of the different baselines:
\begin{equation}
    F_j = \sum_{k\neq{}j}(F_j+F_k)/3 - \sum_{\substack{k, l \neq{}j\\ k<l}}(F_k + F_l)/6  \label{eq:f}
.\end{equation}

In the GRAVITY FT, the 24 outputs (4 outputs per each of the 6 ABCD combiners) are dispersed over 6 wavelength channels in the K-band, and recorded on a SAPHIRA detector \citep{2014SPIE.9148E..17F}. We refer to Figure~1 in Paper~1 for an overview of the pixel arrangement on the detector Real Time Display (RTD). As shown by Equations \ref{eq:a} to \ref{eq:f}, for a given wavelength channel, the column vector $(q_1, \dots{}, q_{24})^T$ containing the 24 different intensity outputs and the column-vector $(F_j, \Gamma_\base{j}{k})^T$ containing the 10 incoherent and coherent flux values are related by a multiplication by a single matrix: the $\mathrm{P2VM}$ (Pixel to Visibility Matrix):
\begin{equation}
  \mathsf{P2VM} \cdot{} (q_1, \dots{}, q_{24})^T = (F_1, \dots{}, F_4, \Gamma_\base{1}{2}, \dots{}, \Gamma_\base{3}{4})^T 
\label{eq:P2VM}
.\end{equation}

The incoherent and coherent fluxes are computed in real time for each of the wavelength channels, which are treated independently. Similarly, if polarisation is split, the two polarisation are treated independently. 

In practice, the ABCD combiners are not perfectly balanced, and the phase offsets can be different from their fiducial values, which means that the coefficients linking the ABCD outputs to the complex amplitudes of the electric field in Equations \ref{eq:a} to \ref{eq:d} can vary. Therefore, the P2VM must be calibrated, which in GRAVITY is done during daytime using an internal source. This P2VM formalism was initially introduced by \citet{2007A&A...464...29T} for the AMBER instrument \citep{2007A&A...464....1P}. It was then adapted to ABCD combiners by \citet{2008SPIE.7013E..16L}, and more details on its implementation can be found in these papers, along with Paper~1.

\subsection{From fluxes to phase/group delays and closures}

After the $\mathsf{P2VM}$ calculation, the resulting wavelength dependent coherent fluxes $\Gamma_{\base{j}{k}, \lambda}$ are still affected by dispersion (both atmospheric dispersion, and dispersion in the fibre delay lines, FDDL), which at first order introduces a phase curvature of the form $e^{iD\left(1-\lambda_0/\lambda\right)^2}$, where $\lambda_0=2.2\,\mu$m. This is corrected by using an empirical value for $D$, which depends on the position of the star in the sky, and on the position of the FDDLs. This correction was calibrated during the first year of on-sky observations.

From this dispersion-corrected coherent flux measurement, we calculate the OPD for each baseline, denoted $\phi_\base{j}{k}$, using:
\begin{align}
  \phi_\base{j}{k} & =\frac{\lambda_0}{2\pi} \times \arg\left(  \sum_{\lambda=1}^{N_\lambda}  \Gamma_{j,k,\lambda} \right) \label{eq:pd}
.\end{align}
These values are concatenated to form the OPD vector $\Phi$, using the convention: $\Phi = \left(\phi_{21}, \phi_{31}, \phi_{41}, \phi_{32}, \phi_{42}, \phi_{43}\right)^T$.
The closure phase $\theta_{j,k,l}^{\rm PD}$ over the triangle linking telescopes $T_j, T_k$ and $T_l$ is calculated in unit of length using:
\begin{align}
\theta_{j,k,l}^{\rm PD} & =\frac{\lambda_0}{2\pi} \times \arg \left( \left\langle    \sum_{\lambda=1}^{N_\lambda} \Gamma_{\base{j}{k},\lambda}  \sum_{\lambda=1}^{N_\lambda} \Gamma_{\base{k}{l},\lambda}  \sum_{\lambda=1}^{N_\lambda} \Gamma_{\base{j}{l},\lambda}^* \right\rangle_{\rm 350 DITs} \right) \label{eq:cp}
,\end{align}
\noindent{}where the incoherent flux is first averaged over 350 Detector Integration Times (DITs, i.e. bout 385~ms at 909~Hz) to boost signal-to-noise, the phase closure being a notoriously noisy quantity.

To calculate the group delay $\psi_\base{j}{k}$ and group closure $\theta_{j,k,l}^\mathrm{GD}$, we first correct the coherent flux for any phase offset constant in wavelength by subtracting the phase delay, and we then average it over 150 DITs (about 165~ms at 909~Hz), again to boost signal-to-noise ratio:
\begin{equation}
  \Gamma_{\base{j}{k},\lambda}'= \left\langle \Gamma_{\base{j}{k},\lambda} \exp\left[-i\frac{2\pi}{\lambda_0}\phi_{j, k}\right] \right\rangle_{\text{150 DITs}}
  \label{eq:gd1}
.\end{equation}The group delay is then given by:
\begin{equation}
  \psi_{j,k} =\frac{\Delta \lambda}{2\pi} \times \arg\left(  \sum_{\lambda=1}^{N_\lambda-1}   \Gamma_{\base{j}{k},\lambda+1}'    \Gamma_{\base{j}{k},\lambda}'^* \right)
\label{eq:gd2}
,\end{equation}
where $\Delta\lambda$ is the  difference between the effective wavelength of two pixel bins. From this, we also construct another vector in OPD-space: $\Psi = \left(\psi_{21}, \psi_{31}, \psi_{41}, \psi_{32}, \psi_{42}, \psi_{43}\right)^T$. 
The closure of the group delay can be written as:
\begin{small}
\begin{equation}
  \theta_{j,k,l}^{\rm GD}= \frac{\Delta \lambda}{2\pi} \times{} \arg \left( \left\langle \sum_{\lambda=1}^{N_\lambda-1}  \Gamma_{\substack{\base{j}{k},\\\lambda+1}}' \Gamma_{\substack{\base{j}{k},\\\lambda}}'^* \sum_{\lambda=1}^{N_\lambda-1} \Gamma_{\substack{\base{k}{l},\\\lambda+1}}' \Gamma_{\substack{\base{k}{l},\\\lambda}}'^*  \sum_{\lambda=1}^{N_\lambda-1} \Gamma_{\substack{\base{j}{l},\\\lambda+1}}'^* \Gamma_{\substack{\base{j}{l},\\\lambda}}' \right\rangle_{\text{350 DITs}}\right)
.\end{equation}
\end{small}

The last quantity of interest for the control algorithm is the uncertainty on the estimation of the OPD, $\sigma_{\base{j}{k}}$. This uncertainty can be computed by first introducing the interferometric SNR, defined as the ratio between the modulus of the coherent flux and the square-root of its variance. We approximate this variance as the half sum of the variance of the real and imaginary parts of the coherent flux, and we average the SNR over 3 DITs\footnote{\textcolor{black}{We thank the referee for pointing out that calculating this SNR as the ratio between the average of the numerator and the average of the denominator (as opposed to the average of the ratio) would likely result in a better estimate. We plan on testing this and implementing it in a future version of the FT.}}:
\begin{equation}
\label{eq:snr_interf}
\mathrm{SNR}_{j,k} = \left \langle \frac{ \left| \sum_\lambda \Gamma_{\base{j}{k},\lambda} \right|}{\sqrt{\frac{1}{2}\sum_\lambda \mathrm{Var}(\Re\Gamma_{\base{j}{k},\lambda}) + \frac{1}{2}\sum_\lambda \mathrm{Var}(\Im\Gamma_{\base{j}{k},\lambda})}} \right\rangle_{\text{3 DITs}}
.\end{equation}
\noindent{}The uncertainty on the phase can then be computed using a similar approach as in the Appendix of \cite{Shao1988}:
\begin{equation}
\frac{1}{\sigma_{\base{j}{k}}} = \frac{2\pi}{\lambda_0}\times{}\mathrm{SNR}_{j,k}
\label{eq:sigma}
.\end{equation}\noindent{}In Equation~\ref{eq:snr_interf}, the variance of the real and imaginary parts of the coherent flux are estimated for each frame from the background, detector and photon noises. Using these OPD noise estimators, and under the assumption that each baseline noise is uncorrelated, we also write the covariance matrix \(W\) on $\Phi$:
\begin{equation}
W=\text{diag}\left(\sigma_{21}^2, \sigma_{31}^2, \sigma_{41}^2, \sigma_{32}^2, \sigma_{42}^2, \sigma_{43}^2\right).
\label{eq:W}
\end{equation}
The variance of $\Gamma$ is estimated from the variance on the individual pixels $q_j$, multiplied by the $\mathsf{P2VM}$ and $\mathsf{P2VM}^T$ matrix, as described in Section 3.4 of Paper~1.
Compared to Paper~1, where the $\sigma_{\base{j}{k}}$ were computed coherently on 5 DITs, the new uncertainties are now computed on individual DITs, and then averaged only over 3 DITs. This change is motivated by the use of an adaptive gain in the new control algorithm, which strongly benefits from a better estimate of the instantaneous signal-to-noise ratio. 

\section{Control loop}
\label{sec:control_loop}

\subsection{The 3 challenges of building a fringe tracker}
\label{sec:challenges}

A first challenge that need to be solved in the implementation of a fringe tracker is that it the dichotomy between two distinct spaces. The FT uses a set of four piezoelectric delay lines \citep{2014SPIE.9146E..23P} to change the OPL of the 4 beams (one per telescope). But the feedback is provided by measurements of the coherent fluxes (see Section~\ref{sec:observables}), which are transformed in a set of 6 measurements of OPD $\phi$ (one per baseline) and group delay $\psi$. The OPL space itself is not observable, and this mismatch between the control space and the measurement space creates a paradox which needs to be solved by the system.

A second challenge arises from the need to regularly update the model used to represent and predict the disturbance. To do so, we use a set of AR models, whose parameters are updated every few seconds via a model fitting routine performed on a separate computer (see hardware description in section~\ref{sec:hardware}). But in order to perform the model fitting, this computer needs to be fed with measurements that represents the disturbance only, excluding the additional piezo command. The control must therefore work with a state model which explicitly separates the OPL in two components: a disturbance component $L$ which represents both atmospheric and ``mechanical'' turbulence, and the OPL introduced by the piezo actuators $X$. Of course, only the total OPD, which derives from the total OPL $L+X$, is actually observable by the fringe-tracker. But in order to be able to update the disturbance model, the two components must be kept separated.

A third challenge comes from the fact that the main phase control loop is fed by phase measurements, which are only known to a modulo $2\pi$ (or $\lambda_0$, in terms of OPD). Thus, this loop is blind to any potential ``fringe shift'', which is the reason why we need the additional group delay loop. In Paper~1, the controller consisted of two completely independent loops running in parallel, with the potential to issue conflicting commands. The algorithm developed at the time did ensure that the integrator of the two controllers would not diverge too widely, but it did not strictly constrain the absolute group delay. The consequence was a tendency of the fringe tracker to jump between fringes during a single observation. The updated architecture detailed in this work provides a much more unified approach to controlling both the group and phase delays, thereby avoiding conflicting commands, mitigating phase jumps and maintaining tracking of the white light fringe.

\subsection{State model}

To solve these challenges, we adopt a state model in OPL space which explicitly separates the disturbance and the OPL introduced by the actuators. We use a set of two state vectors, which both have the unit of length, and which we denote $X_n$ and $L_n$. For simplicity, these state-vectors can be broken down into four independent vectors, corresponding to each telescope. For a given telescope $T_k$, the vector of the last N values of OPL introduced respectively by the piezo actuators ($X_{n, k}$) and by the disturbance ($L_{n, k}$) can be expressed by:
\begin{align}
  X_{n, k} &= \left[x_{k}(t_n), x_{k}(t_{n-1}), \dots, x_{k}(t_{n-N+1})\right]^T ,\\
  L_{n, k} &= \left[l_{k}(t_n), l_{k}(t_{n-1}), \dots, l_{k}(t_{n-N+1})\right]^T
,\end{align}
\noindent{}where $x_k(t_j)$ and $l_k(t_j)$ respectively denotes the OPL introduced by the piezo actuator and disturbance on telescope $T_k$ at time $t_j$. These vectors have a total length of 150, which matches the smoothing length of the group delay. The four telescopes can be concatenated to give our two final state vectors $X_n$ and $L_n$ of dimension $4\times{}150 = 600$:
\begin{align}
X_{n} =  \left[\begin{matrix}
    X_{n, 1}\\
    X_{n, 2}\\
    X_{n, 3}\\
    X_{n, 4}
  \end{matrix}\right]
\quad
L_{n} =  \left[\begin{matrix}
    L_{n, 1}\\
    L_{n, 2}\\
    L_{n, 3}\\
    L_{n, 4}
  \end{matrix}\right]
\label{eq:definitionL}
\end{align}

\begin{figure}
  \centering
  \includegraphics[width=0.49\textwidth]{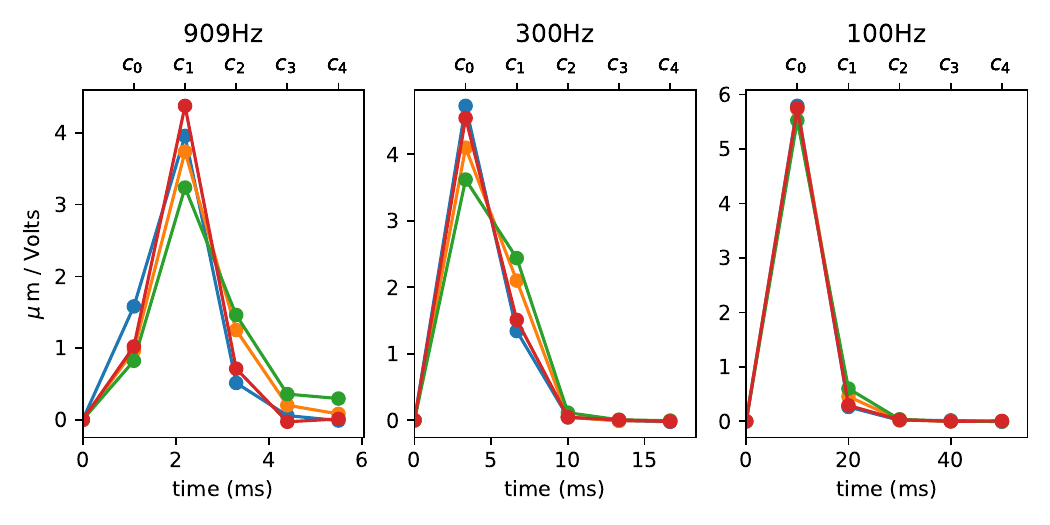}
  \caption{ 
    Open loop response of the piezo control chain. This corresponds to the response of the piezo in micrometers per volt, as a function of time (bottom axis), when submitted to an impulse command. Equivalently, this also gives the values of the coefficients of the fifth order model used in Equation~\ref{eq:piezo} (top axis). Each colour represents a distinct actuator.
    }
   \label{fig:piezo_response}%
\end{figure}

Each of the piezo  control chain (which include the actuator response as well as the pure delay introduced by the processing) has a response function which we model as a fifth-order, where the position of the piezo at time $t_n$ is derived from the command sent at the last five time steps:
\begin{equation}
  x_k(t_{n+1}) = c_{k, 0} u_k(t_{n}) + c_{k, 1} u_k(t_{n-1}) + \dots{} + c_{k, 4} u_k(t_{n-4})
  \label{eq:piezo}
\end{equation}
\noindent{}The $c$ coefficients can be empirically measured by sending an impulse command (in Volts) through the control chain and measuring how the response evolves with time (Figure~\ref{fig:piezo_response}).

We then introduce the command vectors $U_{n, k} = \left[u_{k}(t_n), \dots{}, u_{k}(t_{n-4})\right]$ such that:
\begin{equation}
  x_{k}(t_{n+1}) = \left[\begin{matrix} c_{k,0} & c_{k,1} & c_{k, 2} & c_{k, 3} & c_{k, 4}\end{matrix}\right] \cdot U_{n, k}
.\end{equation} Again, the four telescopes can be concatenated to give $U_n$ of dimension $4\times{}5 = 20$:
\begin{align}
U_{n} =  \left[\begin{matrix}
    U_{n, 1}\\
    U_{n, 2}\\
    U_{n, 3}\\
    U_{n, 4}
  \end{matrix}\right] 
.\end{align}

With these notations, the propagation of the piezo state vector $X_n$ takes the form:
\begin{equation}
  X_{n+1} = A_X \cdot{} X_n + C \cdot{} U_n
  \label{eq:Xpropagation}  
,\end{equation}
\noindent{}where $A_X$ is a $600\times{}600$ block-diagonal matrix that just ``shifts'' the $x$s for each telescope:
\begin{equation}
 A_X = \mathrm{diag}(A, A, A, A)
,\end{equation}with
\begin{equation}
A = \begin{bmatrix}
    0 & 0 & \dots & 0 & 0 \\
    1 & 0 & \dots & 0 & 0 \\
    0 & 1 & \ddots & \vdots & \vdots \\
    \vdots & \ddots & \ddots & 0 & 0 \\
    0 & \dots & 0 & 1 & 0 \\
    \end{bmatrix}
,\end{equation}
\noindent{}and $C$ is also a block-diagonal matrix, but of size $600\times{}20$, which contains the coefficients of the piezo responses:
\begin{equation}
 C = \mathrm{diag}(C_1, C_2, C_3, C_4)
,\end{equation}with
\begin{equation}
  C_k = \begin{bmatrix}
    c_{k, 0} & c_{k, 1} & c_{k, 2} & c_{k, 3} & c_{k, 4} \\
    0 & 0 & 0 & 0 & 0 \\
    \vdots & \vdots & \vdots & \vdots{} & \vdots{} \\
    0 & 0 & 0 & 0 & 0 \\            
  \end{bmatrix}
.\end{equation}

For the disturbance state vector $L_n$, we use a model similar to what was already in use on the GRAVITY FT (see Paper~1), which gave satisfactory results, and which consists in a set of 6 Auto-Regressive (AR) models in OPD space (see Appendix~\ref{sec:AQ}). Since the upgraded version of the FT is now using a state-model in OPL rather than in OPD space, this AR model needs to be converted to OPL space. To do so, we first note that for a 4-telescope (6-baseline) interferometer like GRAVITY, an OPD vector \(\Phi = (\phi_\base{1}{2}, \phi_\base{1}{3}, \phi_\base{1}{4}, \phi_\base{2}{3}, \phi_\base{2}{4}, \phi_\base{3}{4})^T\) and an OPL vector \((l_1, l_2, l_3, l_4)^T\) are linked by the matrix equation:
\begin{equation}
  \Phi = M\cdot{}l,
\end{equation}
where \(M\) is a \(6 \times 4\) matrix defined as:
\begin{align}
  \label{eq:M}
  M = \begin{bmatrix}
      1 & -1 & 0 & 0 \\
      1 & 0 & -1 & 0 \\
      1 & 0 & 0 & -1 \\
      0 & 1 & -1 & 0 \\
      0 & 1 & 0 & -1 \\
      0 & 0 & 1 & -1
    \end{bmatrix}.
\end{align}
\noindent{}Therefore, computing an OPL vector from knowledge of the OPD vector is under-constrained, yielding multiple solutions. A solution entails using the Moore-Penrose pseudo-inverse of \(M\):
\begin{equation}
  l = M^\pinv \cdot{} \Phi.
\end{equation}
For reference, the value of \(M^\pinv\) is:
\begin{align}
\label{eq:Mpinv}
  M^\pinv = \frac{1}{4} \begin{bmatrix}
      1 & 1 & 1 & 0 & 0 & 0 \\
      -1 & 0 & 0 & 1 & 1 & 0 \\
      0 & -1 & 0 & -1 & 0 & 1 \\
      0 & 0 & -1 & 0 & -1 & -1
  \end{bmatrix}.
\end{align}
\noindent{}This particular choice ensures that the mean value of the OPLs on each telescope is $0$, which is beneficial for a fringe-tracker. It helps avoid ``runaway'' situations where all the actuators move toward a large offset, saturating the system.

These equations can be extended to our OPL state-space vector $L_n$ as defined in Equation~\ref{eq:definitionL}, and a similarly defined OPD vector $\Phi_{n} =  (\Phi_{n,\base{1}{2}}, \dots{}, \Phi_{n,\base{3}{4}})^T$ using similar matrices in which the ``1s'' are replaced by identity matrices of size 150:
\begin{align}
  \mathcal{M} &= \left[\begin{matrix}
      I_{150} & -I_{150} & 0 & 0 \\
      I_{150} & 0 & -I_{150} & 0 \\
      I_{150} & 0 & 0 & -I_{150} \\
      0 & I_{150} & -I_{150} & 0 \\
      0 & I_{150} & 0 & -I_{150} \\
      0 & 0 & I_{150} & -I_{150}
    \end{matrix}\right],\\
  \mathcal{M}^{\pinv} &= \frac{1}{4}\left[\begin{matrix}
      I_{150} & I_{150} & I_{150} & 0 & 0 & 0 \\
      -I_{150} & 0 & 0 & I_{150} & I_{150} & 0 \\
      0 & -I_{150} & 0 & -I_{150} & 0 & I_{150} \\
      0 & 0 & -I_{150} & 0 & -I_{150} & -I_{150}
      \end{matrix}\right].
\end{align}
\noindent{}With these matrices, the relationships remain:
\begin{align}
  \Phi_{n} &= \mathcal{M} \cdot{} L_{n} ,\\
  L_n &= \mathcal{M}^{\pinv}\cdot{}\Phi_n  
.\end{align}

Our AR model fitting provides a set of 6 matrices $A_\base{j}{k}$ which can be used to propagate the individual $\Phi_{n, \base{j}{k}}$. These can be combined in a block-diagonal matrix to propagate $\Phi_n$ itself:
\begin{align}
  \Phi_{n+1} = \left[\begin{array}{cccc}
      A_\base{1}{2} & 0 & \dots & 0 \\
      0 & A_\base{1}{3} & \ddots & \vdots \\
      \vdots & \ddots & \ddots & 0 \\
      0 & \dots & 0 & A_\base{3}{4}
    \end{array}\right] \cdot{} \Phi_n
.\end{align}
\noindent{}Switching to OPL space, we have:
\begin{align}
\label{eq:ML=AML}
  \mathcal{M} \cdot{} L_{n+1} = \left[\begin{array}{cccc}
      A_\base{1}{2} & 0 & \dots & 0 \\
      0 & A_\base{1}{3} & \ddots & \vdots \\
      \vdots & \ddots & \ddots & 0 \\
      0 & \dots & 0 & A_\base{3}{4} 
    \end{array}\right] \cdot{} \mathcal{M} \cdot{} L_n
.\end{align}
\noindent{}At this point, we can note that:
\begin{equation}
\mathcal{M}^\pinv \cdot \mathcal{M} =
   I_{600} - 
   \frac{1}{4}
   \begin{bmatrix}
       I_{150} &  I_{150} &  I_{150} &  I_{150} \\
       I_{150} &  I_{150} &  I_{150} &  I_{150} \\
       I_{150} &  I_{150} &  I_{150} &  I_{150} \\
       I_{150} &  I_{150} &  I_{150} &  I_{150}
   \end{bmatrix}
,\end{equation}
\noindent{}which means that as long as our OPL vectors $L_{n+1}$ contain values where the average on all telescopes is 0, we can consider that $\mathcal{M}\cdot{}\mathcal{M}$ is the identity matrix. In other words, as long as:
\begin{equation}
\label{eq:L0mean}
    \begin{bmatrix}
       I_{150} &  I_{150} &  I_{150} &  I_{150} \\
       I_{150} &  I_{150} &  I_{150} &  I_{150} \\
       I_{150} &  I_{150} &  I_{150} &  I_{150} \\
       I_{150} &  I_{150} &  I_{150} &  I_{150}
   \end{bmatrix}
   \cdot{} L_{n+1} = 0
,\end{equation}
\noindent{}we have:
\begin{equation}
\mathcal{M}^\pinv \cdot{} \mathcal{M} \cdot{} L_{n+1} = L_{n+1}.
\end{equation}
\noindent{}Assuming temporarily that $L_{n+1}$ fulfil the condition given by Equation~\ref{eq:L0mean}, we can left-multiply Equation~\ref{eq:ML=AML} by $\mathcal{M}^\pinv$ to get:
\begin{align}
  L_{n+1} = \left( \mathcal{M}^\pinv \cdot{} \left[\begin{array}{cccc}
      A_\base{1}{2} & 0 & \dots & 0 \\
      0 & A_\base{1}{3} & \ddots & \vdots \\
      \vdots & \ddots & \ddots & 0 \\
      0 & \dots & 0 & A_\base{3}{4}
    \end{array}\right] \cdot{} \mathcal{M} \right) \cdot{} L_n
.\end{align}
\noindent{}A block matrix multiplication inside the brackets gives the final form of our model in OPL state-space:
\begin{equation}
  \label{eq:Lpropagation}
  \hat{L}_{n+1} = A_L \cdot L_n
,\end{equation}
\noindent{}with:
\begin{align}
  \setlength\arraycolsep{2pt}
A_L =   \begin{bmatrix}
    \frac{A_\base{1}{2}+A_\base{1}{3}+A_\base{1}{4}}{4} & -A_\base{1}{2}/4 & -A_\base{1}{3}/4 & -A_\base{1}{4}/4 \\
    -A_\base{1}{2}/4 & \frac{A_\base{1}{2}+A_\base{2}{3}+A_\base{2}{4}}{4} & -A_\base{2}{3}/4 & -A_\base{2}{4}/4 \\
    -A_\base{1}{3}/4 & -A_\base{2}{3}/4 & \frac{A_\base{1}{3}+A_\base{2}{3}+A_\base{3}{4}}{4} & -A_\base{3}{4}/4 \\
    -A_\base{1}{4}/4 & -A_\base{2}{4}/4 & -A_\base{3}{4}/4 & \frac{A_\base{1}{4}+A_\base{2}{4}+A_\base{3}{4}}{4}
\end{bmatrix}
\label{eq:A_L}
.\end{align}

We can formulate two important remarks on the propagation model given by Equation~\ref{eq:Lpropagation}. Firstly, since:
\begin{align}
    \begin{bmatrix}
       I_{150} &  I_{150} &  I_{150} &  I_{150} \\
       I_{150} &  I_{150} &  I_{150} &  I_{150} \\
       I_{150} &  I_{150} &  I_{150} &  I_{150} \\
       I_{150} &  I_{150} &  I_{150} &  I_{150}
   \end{bmatrix}
   \cdot{} A_L = 0
,\end{align}
our propagation model always guarantees that the condition given by Equation~\ref{eq:L0mean} is fulfilled. Physically, this stems from the fact that our propagation model is really the conversion of an OPD model to OPL space, where we used the freedom given by the under-constrained nature of the conversion to explicitly guarantee that the average piston is always 0, through the definition of $M^\pinv$ in Equation~\ref{eq:Mpinv}. Secondly, the presence of $\hat{L}$ with a ``hat'' in Equation~\ref{eq:Lpropagation} emphasises the fact that this is only a model-propagated estimate of $L$. There is no hat for $X$ in Equation~\ref{eq:Xpropagation}, since we consider our piezo model to be perfect (or at least much better than our atmospheric model).

Equation~\ref{eq:Lpropagation} is linear, and the covariance matrix on \(L_n\) is propagated accordingly, using an additional \(600 \times 600\) matrix \(Q\), which represents the ``process noise''.
\begin{equation}
  \hat{P}_{n+1} = A_L \cdot P_n \cdot {A_L}^T + Q.
  \label{eq:Ppropagation}
\end{equation}
The process noise \(Q\) is derived from a set of \(150 \times 150\) matrices \(Q_{j,k}\) using a similar block construction:
\begin{align}
  \setlength\arraycolsep{2pt}
Q =   \begin{bmatrix}
    \frac{Q_\base{1}{2}+Q_\base{1}{3}+Q_\base{1}{4}}{4} & -Q_\base{1}{2}/4 & -Q_\base{1}{3}/4 & -Q_\base{1}{4}/4 \\
    -Q_\base{1}{2}/4 & \frac{Q_\base{1}{2}+Q_\base{2}{3}+Q_\base{2}{4}}{4} & -Q_\base{2}{3}/4 & -Q_\base{2}{4}/4 \\
    -Q_\base{1}{3}/4 & -Q_\base{2}{3}/4 & \frac{Q_\base{1}{3}+Q_\base{2}{3}+Q_\base{3}{4}}{4} & -Q_\base{3}{4}/4 \\
    -Q_\base{1}{4}/4 & -Q_\base{2}{4}/4 & -Q_\base{3}{4}/4 & \frac{Q_\base{1}{4}+Q_\base{2}{4}+Q_\base{3}{4}}{4}
\end{bmatrix}.
\label{eq:Q}
\end{align}

The matrices \(A_{j,k}\) and \(Q_{j,k}\) are recalculated every 5 seconds by an asynchronous machine (\texttt{wgvkalm} described in sec~\ref{sec:hardware}) based on OPD measurements. This computation makes use of the Python toolbox "Time Series Analysis" \citep{seabold2010statsmodels}. A detailed explanation of its implementation is provided in Appendix~\ref{sec:AQ}.

\subsection{Phase control}
\label{sc:K}

Without any ``feedback'', i.e. any additional information about the state of the system, the internal model of the FT would simply keep propagating the state-vector $L$ indefinitely using Equation~\ref{eq:Lpropagation}, and dispatch commands to the piezo actuators to respond to its own predictions. This behaviour would probably be acceptable over a few iterations, but given the uncertainties on the predictions of the disturbance, the internal state-vector $L_n$ would quickly diverge from the real values. The internal model is ``aware'' of this, as Equation~\ref{eq:Ppropagation} shows that in such a scenario, the covariance matrix would slowly inflate due to the accumulating ``process noise'' $Q$.

To properly track the fringes over long periods, the FT therefore needs to integrate measurements into the loop. But these measurements should not be taken at face value either, as they are also affected by varying levels of uncertainties. Integrating noisy measurements into the loop without careful consideration will create additional noise in the system, thereby reducing its overall performance. The main task of the Kalman controller in phase-delay loop is to determine how to best balance the information coming from the state model propagation, and from the additional measurements performed on the system. In Paper~1, this was done using a fixed asymptotic Kalman gain, which could not react in real time to the quality of the model and measurements. This new version of the Kalman provides an optimal gain, calculated at each iteration to combine the two sources of information in the best possible way. 

At each time step $t_n$, we obtain a set of 6 measurements of the phase (or OPD) on each baseline, with an estimate of their error bars. The measurement process is detailed in Section~\ref{sec:observables}, and for the context of this Section, we will simply denote $\Phi_n = (\phi_\base{j}{k}(t_n))$ the column vector of size 6 containing the individual phase measurements (see Equation~\ref{eq:pd}), and $W_n$ the associated $6\times{}6$ diagonal covariance matrix, as defined from Equation~\ref{eq:sigma} in Equation~\ref{eq:W}.

These measurements are related to our state-vectors $L_n$ and $X_n$ through the ``measurement equations'':
\begin{equation}
  \phi_\base{j}{k}(t_n) = \left[l_k(t_n) - l_j(t_n) + x_k(t_n) - x_j(t_n)\right]_{\text{mod }\lambda_0}
.\end{equation}
\noindent{}This means that although our state-vectors are of length 600 to accommodate for the group-delay measurements which are averaged over 150 DITs, the OPD measurement itself only relies on the first values (at time $t_n$) for each telescope. In matrix form, this can be written:
\begin{align}
  \Phi_{n} = \left[H_{\phi} \cdot{} (L_n+X_n)\right]_{\text{mod }\lambda_0}
,\end{align}
\noindent{}with $H_\phi$ defined by
\begin{align}  
\label{eq:Hphi}
  H_{\phi} = \begin{bmatrix}
    \boldsymbol{v}_{1} & -\boldsymbol{v}_{1} & 0 & 0 \\
    \boldsymbol{v}_{1} & 0 & -\boldsymbol{v}_{1} & 0 \\
    \boldsymbol{v}_{1} & 0 & 0 & -\boldsymbol{v}_{1} \\
    0 & \boldsymbol{v}_{1} & -\boldsymbol{v}_{1} & 0 \\
    0 & \boldsymbol{v}_{1} & 0 & -\boldsymbol{v}_{1} \\
    0 & 0 & \boldsymbol{v}_{1} & -\boldsymbol{v}_{1}
  \end{bmatrix}
,\end{align}
\noindent{}where $\boldsymbol{v}_1 = (1, 0, \dots{}, 0)^T$ is the first basis vector of length $150$.

Given our assumption that the piezo state-vector $X_n$ is perfectly known, the difference between the actual measurement and the expected measurement $\hat{\Phi}_n = H_\phi \cdot (\hat{L}_n+X_n)$ is directly related to the difference between the actual state of the disturbance $L_n$ and its estimate $\hat{L}_n$:
\begin{equation}
  \Phi_n - \hat{\Phi}_n = \left[H_\phi \cdot \left(L_n - \hat{L}_n\right)\right]_{\text{mod }\lambda_0}
  \label{eq:phi_error}
\end{equation}
\noindent{}and so, ignoring the modulo, it is tempting to simply update the state-vector according to $L_n = \hat{L}_n + {H_\phi}^\pinv \cdot (\hat{\Phi}_n - \Phi)$. But this would be reinjecting all the noise from the measurement into the control loop, without any consideration for the respective levels of confidence of the said measurement and the prediction from the model propagation.

These levels of confidence are represented by the matrices $\hat{P}_n$ and $W_n$, which can therefore be used to calculate a Kalman gain which properly weights the model prediction and the measurement \cite[section 4.3, for a proper derivation of this gain]{Nowak2019}:
\begin{align}
\label{eq:Kgain}
  K_{n} = \left(\hat{P}_{n} \cdot{} {H_{\phi}}^T\right)\cdot{}\left( H_{\phi} \cdot{} \hat{P}_{n} \cdot{} {H_\phi}^T + W_n \right)^{-1}
.\end{align}
\noindent{}The update state-vector and its covariance matrix are then given by: 
\begin{align}  
  L_{n} &= \hat{L}_{n} + K_n \cdot{} \left[ \hat{\Phi}_{n} - \Phi_{n} + \Phi_{{\rm CP},n}\right]_{\text{mod }\lambda_0} \label{eq:update}, \\
  P_{n} & = \hat{P}_n + K_n \cdot{} (H_\phi \cdot{} \hat{P}_n \cdot{} {H_\phi}^T + W_n) \cdot{} {K_n}^T
  \label{eq:Pkalman}
,\end{align}
\noindent{}which does indeed reduce to $L_n = \hat{L}_n + {H_\phi}^{\pinv} \cdot (\hat{\Phi}_n - \Phi_n + \Phi_{\mathrm{CP}, n})$ in the $W_n \ll \hat{P}_n$ regime, and also to $L_n = \hat{L}_n$ in the $\hat{P}_n \ll W_n$ regime. 

The introduction of $\Phi_\mathrm{CP}$ in this equation ensures a zero closure-phase on the measurement error. It is crucial to understand that the model being in OPL space, it cannot account for any non-zero closure phase. This closure-phase issue was already present in the first version of the FT, and we refer to Section 4.2 of Paper~1 for a detailed explanation. We simply recall here that $\Phi_\mathrm{CP}$ is constructed by setting three of the four closure phases $\theta_{j,k,l}^{\rm PD}$ to the appropriate baselines, ignoring the closure phase on the triangle of lowest signal-to-noise ratio:
\begin{equation}
\Phi_{\rm CP}=
  \left( \begin{array}{c} \theta_{1,2,4}^{\rm PD}\\ \theta_{1,3,4}^{\rm PD}\\ 0\\ \theta_{2,3,4}^{\rm PD}\\ 0\\ 0  \end{array}\right)  
 \text{or} \left( \begin{array}{c}  \theta_{1,2,3}^{\rm PD}\\ 0\\ -\theta_{1,3,4}^{\rm PD}\\ 0\\ -\theta_{2,3,4}^{\rm PD}\\ 0  \end{array}\right)  
 \text{or} \left( \begin{array}{c} 0\\ -\theta_{1,2,3}^{\rm PD}\\ -\theta_{1,2,4}^{\rm PD}\\ 0\\ 0\\ \theta_{2,3,4}^{\rm PD}  \end{array}\right)
 \text{or} \left( \begin{array}{c}  0\\ 0\\ 0\\ \theta_{1,2,3}^{\rm PD}\\ \theta_{1,2,4}^{\rm PD}\\ \theta_{1,3,4}^{\rm PD} \end{array}\right)  
 \label{eq:RefPD}
.\end{equation}

It is also important to note that given such a definition, two fundamentally different kinds of variations of $\Phi_\mathrm{CP}$ can occur: (1) $\Phi_\mathrm{CP}$ can change because of slowly varying closures phases in the measurements, in which case it is simply tracking actual variations; (2) $\Phi_\mathrm{CP}$ can switch between options given in Equation~\ref{eq:RefPD}, which occurs when the triangle of lowest signal to noise ratio changes. The second case corresponds to a change of configuration, leading to an abrupt variation of $\Phi_\mathrm{CP}$, and is therefore banned while the science detector is integrating. This behaviour is controlled thanks to an internal boolean parameter, called the mobility flag, which is "False" during science integrations.

\subsection{Group delay control}
\label{sc:gd}

The initial version of the FT described in Paper~1 was designed for an instrument performing prolonged exposures at a consistent wavelength of $2.2\,\mu\text{m}$. As long as it stayed in the coherence envelope of the fringes, the FT had the freedom to transition from one fringe to another. This is unacceptable if the GRAVITY fringe tracker is to be used to feed a different science instrument operating at a different wavelength, as this would blur the fringes on the science detector. In an effort to make the GRAVITY FT compatible with the L-band combiner MATISSE \citep{2022A&A...659A.192L}, the group delay controller has been modified to consistently track the same  fringe during a scientific exposure. This new MATISSE mode is called GRA4MAT \citep{Woillez+2023}, and is now routinely offered to the community.

For this, the group delay control loop must determine in real time which fringe is being tracked by the phase-control loop, and updates the state-vector $L_n$ by integer multiple of $\lambda_0$ if required, depending on which fringe it wants to track. Such changes are completely transparent to the phase control loop, due the modulo $\lambda_0$ in Equation~(\ref{eq:phi_error}).

To implement this behaviours, we use a formalism similar to the phase-delay loop. We first introduce the $6\times{}600$ measurement matrix $H_\psi$ to relate the group-delay $\Psi$ to the state-vectors $L_n$ and $X_n$:
\begin{align}
  \Psi_{n} = H_{\psi} \cdot{} (L_n+X_n) \\
,\end{align}
\noindent{}with
\begin{align}
\label{eq:Hpsi}
H_{\psi} = \frac{1}{150}\begin{bmatrix}
    \boldsymbol{1} & -\boldsymbol{1} & 0 & 0 \\
    \boldsymbol{1} & 0 & -\boldsymbol{1} & 0 \\
    \boldsymbol{1} & 0 & 0 & -\boldsymbol{1} \\
    0 & \boldsymbol{1} & -\boldsymbol{1} & 0 \\
    0 & \boldsymbol{1} & 0 & -\boldsymbol{1} \\
    0 & 0 & \boldsymbol{1} & -\boldsymbol{1}
  \end{bmatrix}
,\end{align}
\noindent{}and where $\boldsymbol{1} = (1, 1, \dots{}, 1)$ is a row-vector of length $150$ full of ``1s''. This is similar to the definition of $H_\phi$, but includes an averaging of the past 150 DITs.

We also introduce $\Psi_{\rm CP}$, an analogue of $\Phi_{\rm CP}$ which is calculated using $\theta_{j,k,l}^{\rm GD}$ instead of $\theta_{j, k, l}^\mathrm{PD}$ in Equation~\ref{eq:RefPD}, and an additional vector $\Psi_{\rm Zero GD}$. This additional setpoint is used to select which specific fringe will be set at a group delay of 0, to be tracked by the FT. Again, to avoid fringe jumps during a science integration, a mobility flag is used to freeze $\Psi_{\rm Zero GD}$ during integrations on the science detector. When the mobility flag allows for changes, the $\Psi_\text{Zero GD}$ setpoint is estimated so that:
\begin{equation}
  \left[\Psi_n - \hat{\Psi}_n - \Psi_{\rm CP} - \Psi_{\rm Zero GD} \right]_{\text{mod }lambda_0} = 0\,.
 \label{eq:setmod}
\end{equation}
This condition is valid modulo $\lambda_0$, which means that Equation~\ref{eq:setmod} does not determine exactly which fringe will be tracked.
To lift this uncertainty, two requirements must be added:
\begin{enumerate}
 \item[R1:] Vector $\Psi_{\rm Zero GD}$ must be as small as possible.
\item[R2:] Vector $\Psi_{\rm Zero GD}$ must not have a closure phase.
\end{enumerate}
The first requirement is to ensure that the system is tracking the white light fringe. The second requirement is needed because the closure phase has already been nulled by the $Psi_{\rm CP}$ term.

Therefore, the zero group delay setpoint is estimated by the real time computer as:
\begin{equation}
  \Psi_{\rm Zero GD} = M \cdot \left[ M^\dag \cdot{} (\Psi_n - \hat{\Psi}_n - \Psi_\mathrm{CP}) \right]_{\text{mod }\lambda_0}
,\end{equation}
where $M$ was introduced in Equation~\ref{eq:M}, and $M^\dag$ is a pseudo inverse weighted by the covariance matrix $W_n$, i.e. the matrix that solves the linear system $Y=MX$ in the least-square sense, considering that $W_n$ from by Equation~\ref{eq:W} is the covariance matrix of $X$. This matrix $M^\dag$ can be calculated using:
\begin{equation}
  M^\dag  =  ( M^T \cdot W_n \cdot M)^{-1} \cdot M^T \cdot{} W_n
  \label{eq:Mdag}
.\end{equation}

The controller ensures that the group delay never deviates by more than $\lambda_0/2$ from this setpoint. For this, the controller calculates at each cycle an error in the form of $\Psi - \hat{\Psi} - \Psi_{\rm CP} -  \Psi_\text{Zero GD}$. It converts this group delay error to OPL space using $M^\dag$ from Equation~\ref{eq:Mdag}. If this error exceeds  $\lambda_0/2$ on any telescope, the controller subtracts $\lambda_0$ to all the 150 values corresponding to this telescope in the state-vector $L$. Similarly, if an OPL value falls below $-\lambda/2$, the elements of $L$ corresponding to this telescope are increased by $\lambda_0$. These adjustments by integer multiples of $\lambda_0$ never disturb the phase controller, due to the presence of a modulo $\lambda_0$ in Equation.~\ref{eq:update}, and thanks to the stationarity of the propagation model (more details in Appendix~\ref{sec:AQ}).

\subsection{Commands to the actuators}
\label{sc:F}

The command for the actuators is obtained from:
\begin{equation}
\label{eq:command}
  U_{n+1} = F \cdot  L_{n} + U_{{\rm setpoint },\,n} + U_{{\rm search },\,n}\,.
\end{equation}
\noindent{}In this equation, \(U_{\rm setpoint }\) is used to modulate the position of the fringes, in a process detailed in Paper~1\footnote{\(U_{\rm setpoint }\) is called \(U_{\rm modulation}\) in Paper~1.}. This setpoint can now also be controlled externally by another machine -- for example,
 to correct non-common optical path delays -- as described in Section~\ref{sec:hardware}. 

The term \(U_{\rm search }\) is used to search for the fringes when they are lost by one or more telescopes. This is made to ensure that the FT continues to track the fringes on the baselines where they remain visible, while searching on the others.

The term \(F \cdot  L_{n}\) cancels out the disturbance calculated by the FT. More precisely, this term is used to negate the influence of the disturbance within a given number of ``future'' DITs. This number of DITs, $d$, is made to be adjusted to the mean response speed of the fringe tracker. Practically, it is $d=2$ at 909\,Hz, and $d=1$ at lower frequencies. Hence, $F$ is calculated such as:
\begin{equation}
 C\cdot F \cdot  L_{n} = H_\phi \cdot A^d \cdot L_n \,.
\end{equation}
Which is obtained to the best of our capabilities, by taking:
\begin{equation}
  \setlength\arraycolsep{2pt}
  F= \frac{1}{4} \begin{bmatrix}
    \frac{1}{C_{T_1}} & \frac{1}{C_{T_1}} & \frac{1}{C_{T_1}} & 0 & 0 & 0 \\
    \frac{-1}{C_{T_2}} & 0 & 0 & \frac{1}{C_{T_2}} & \frac{1}{C_{T_2}} & 0 \\
    0 & \frac{-1}{C_{T_3}} & 0 & \frac{-1}{C_{T_3}} & 0 & \frac{1}{C_{T_3}} \\
    0 & 0 & \frac{-1}{C_{T_4}} & 0 & \frac{-1}{C_{T_4}} & \frac{-1}{C_{T_4}} \\
    \end{bmatrix} \cdot  H_\phi \cdot  A^d\,
\end{equation}
with \(C_{T_k}\) representing the integrated impulse response of the piezo actuator at telescope \(k\): 
\begin{equation}
    C_{T_k} = \sum_{n=0}^4 c_{k,n},
\end{equation}
\noindent{}where the $c$ coefficients are the same as in Equation~\ref{eq:piezo}

\subsection{Summary of the new control loop}
\label{sec:summary}

\begin{figure*}
  \centering
  \includegraphics[width=\textwidth]{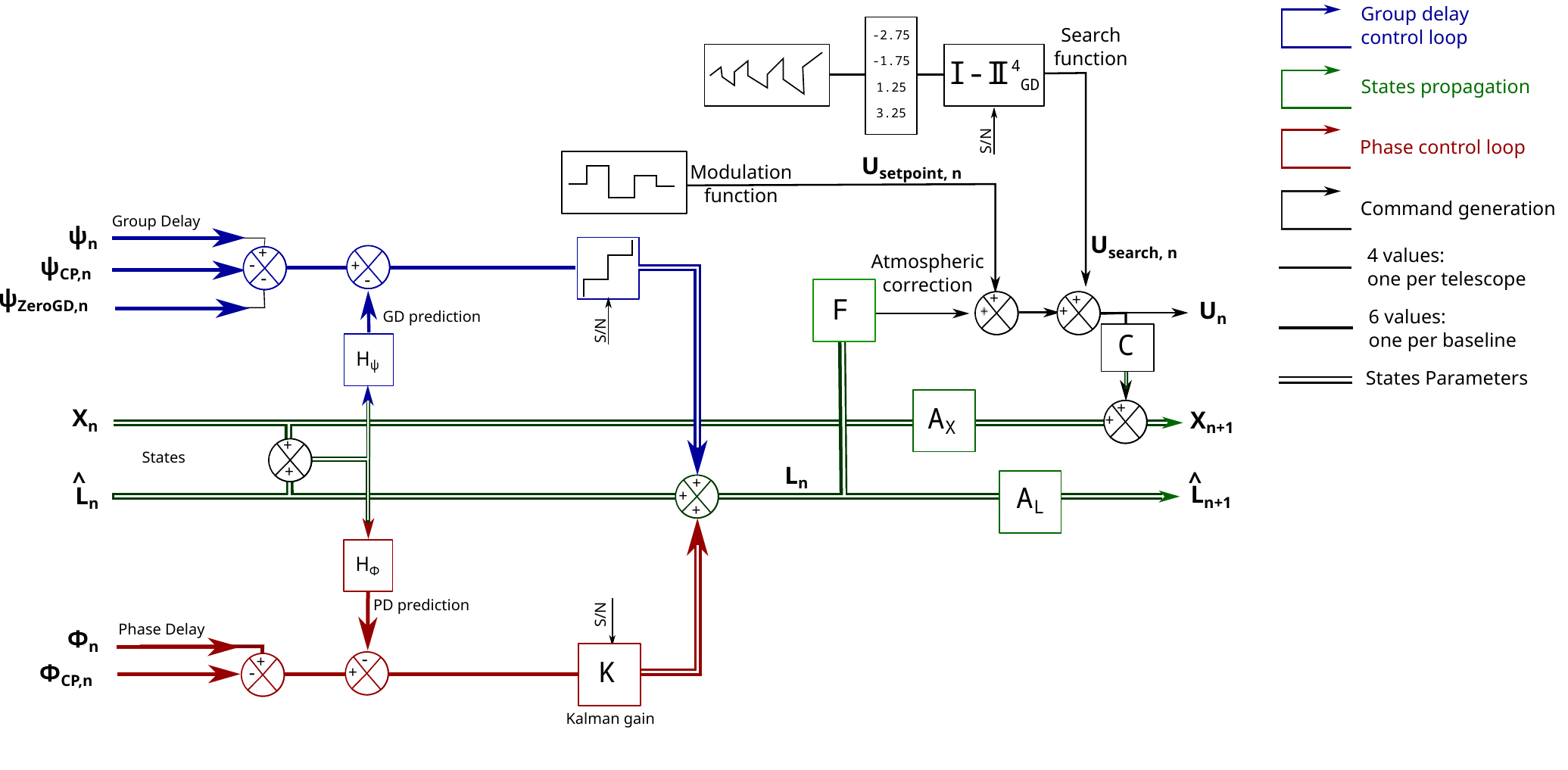}
  \caption{Block diagram of the GRAVITY fringe-tracking controller within a single integration time, denoted as $n$. Central to the fringe tracker are the parameter states, illustrated in green, which represent the disturbance conditions ($L_n$) and actuator positions ($X_n$). The blue segment indicates the group-delay feedback applied to the disturbance state, while the red segment denotes the phase-delay feedback on $L_n$, and the black segments represent the command send to the actuators.}
  \label{fig:control}
\end{figure*}

The architecture of the new control loop is illustrated in Figure~\ref{fig:control}. As mentioned, in this updated architecture, the two control loops do not run in parallel anymore, as it was the case in Paper~1. To some extent, the phase loop can now be seen as the main control loop, and the group-delay control as an auxiliary loop which provides additional update of the state parameters to ensure proper tracking of the group delay. The phase tracking is done using a Kalman filter which does not rely on asymptotic gain anymore, but which uses a real-time covariance-weighted combination of a model prediction (in OPL space) and a measurement update (in OPD space). The control algorithm consists in the following steps:
\begin{enumerate}
  \itemsep0.5em

\item At iteration $n$, we start with an estimate of the state-vector $\hat{L}_n$, propagated from the previous iteration. Our state-vector $L$ contains the path length at each telescope over a given number $N = 150$ previous iterations. This estimate, which also comes with a covariance matrix $\hat{P}_n$, represents our best knowledge of the state parameters. We also have a state-vector $X_n$, which stores the position of the piezo actuators over the last 150 iterations.
  
\item From these state-vectors, we construct the total OPL at each telescope, which is simply $\hat{L}_n + X_n$, and then calculate the expected values for the phase and group delay ($\hat{\Phi}$ and $\hat{\Psi}_n$), using a set of ``measurement'' matrices $H_\phi$ and $H_\psi$, respectively given in Equations~\ref{eq:Hphi} and \ref{eq:Hpsi}, and which convert our OPL-based vectors to our OPD-based measurements:
  \begin{align}
    \hat{\Phi}_n &= H_\phi \cdot (\hat{L}_{n} + X_n) \\
    \hat{\Psi}_n &= H_\psi \cdot (\hat{L}_{n} + X_n)    
  \end{align}

\item A measurement $\Phi$ of the OPDs is performed, which comes with an estimate of its uncertainties, in a process described in Section~\ref{sec:observables}. The difference between this measurement and the expected value $\hat{\Phi}_n$ encodes some information on the state of the disturbance. This information is combined with the a priori knowledge of the state vector $\hat{L}_n$ to give an updated ``a posterori'' estimate $L_n$. This is a form of optimal data-fusion, which uses a ``Kalman gain'' $K_n$, defined in Equation~\ref{eq:Kgain}:
  \begin{equation}
    L_{n} = \hat{L}_{n} + K_{n} \cdot \left[\hat{\Phi}_n - (\Phi_{n} - \Phi_{{\rm CP},n})\right]_{\mathrm{mod}\,\lambda_0}
  \end{equation}
  \noindent{}The use of $\Phi_\mathrm{CP}$ arises from fact that the controller, which works in OPL, cannot account for a non-zero phase closure. At this step, the covariance $P_n$ on $L_n$ is propagated from $\hat{P}_n$, using Equation~\ref{eq:Pkalman}.

\item The measured and expected group delays $\Psi$ and $\hat{\Psi}$ are used to determine whether $L_n$ needs to be further modified by an increment of integer multiples of $\lambda_0$. This is the group-delay tracking, explained in Section~\ref{sc:gd}.

\item The vector $L_n$ contains our best estimate of the disturbance, and is used to generate a command $U_{n+1}$, in a process described by Equation~\ref{eq:command}.
  
\item The state-vectors are propagated according to their respective model:
  \begin{align}
    X_{n+1} &= A_X \cdot X_n + C \cdot U_{n+1}\\
    \hat{L}_{n+1} &= A_L \cdot L_n 
  \end{align}
 The covariance matrix $P_n$ on $L_n$ is also propagated into $\hat{P}_{n+1}$, using:
  \begin{equation}
    \hat{P}_{n+1} = A_L \cdot P_n \cdot {A_L}^T + Q.
  \end{equation}
  \noindent{}After this propagation, the next iteration can start.
\end{enumerate}

\begin{figure}
  \centering
  \includegraphics[width=0.5\textwidth]{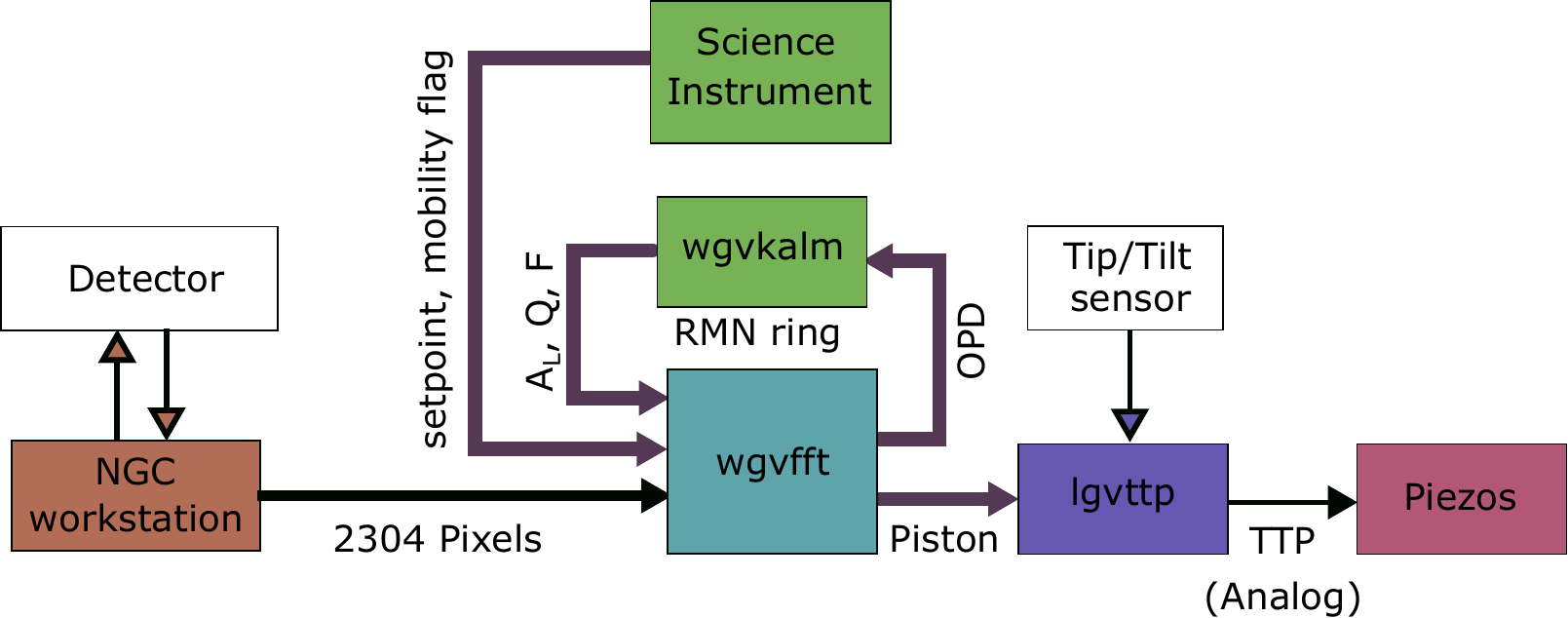}
  \caption{Hardware architecture of the GRAVITY Fringe Tracker. The real-time controller operates within the \texttt{wgvfft} Linux workstation. Another Linux workstation, \texttt{wgvkalm}, is dedicated to calculating the Kalman filter parameters based on the observed OPDs. A Motorola CPU in the \texttt{lgvttp} workstation integrates piston information with tip/tilt sensor data to accurately adjust mirrors mounted on piezoelectric actuators in Tip-Tilt-Piston (TTP). Additionally, an external workstation can provide the fringe tracker with piston setpoints and a mobility flag, influencing the controller’s behaviour.}
  \label{fig:Architecture}
\end{figure}

\begin{figure*}
  \centering
  \includegraphics[width=\textwidth]{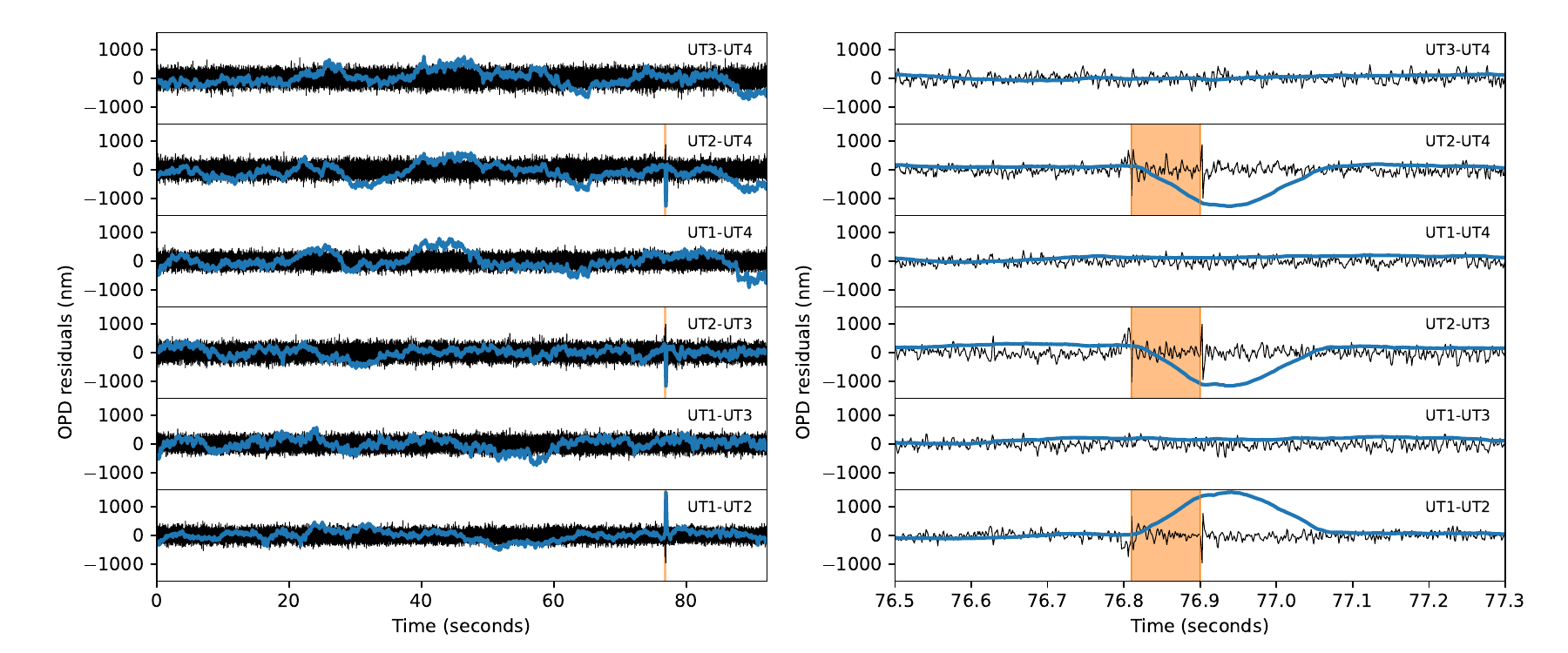}
  \includegraphics[width=\textwidth]{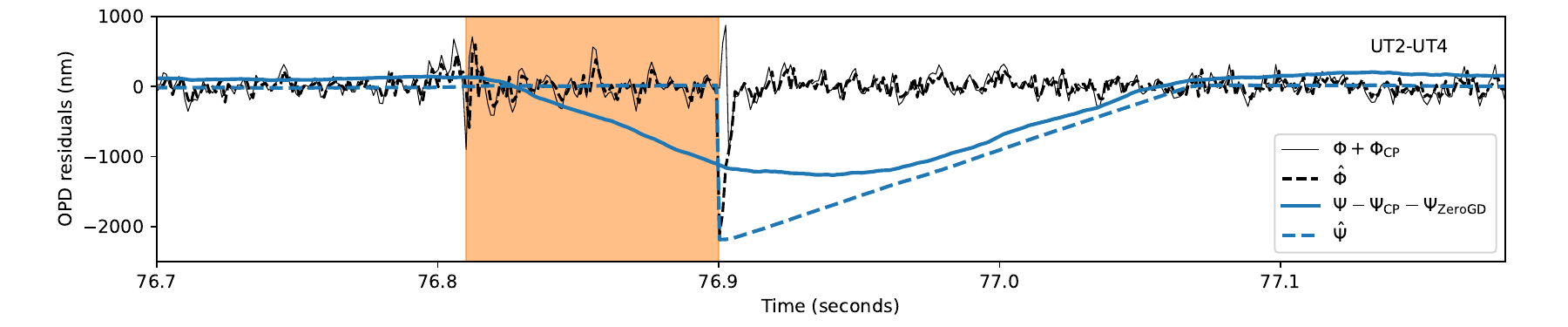}
  \caption{Observation of star HR~8799 using GRAVITY. The data is from a single DIT exposure on the spectrometer, spanning 100 seconds. Black curves denote the OPD, derived from the phase delay: $\left(\Phi - \Phi_{\rm CP}\right)$. Blue curves illustrate observations from the group delay: $\left(\Psi - \Psi_{\rm CP} - \Psi_{\rm Zero GD}\right)$. Each panel represents a different baseline. At $t_1=76.81$ seconds, a noticeable fringe jump occurred on UT2. The effect of this jump across all baselines is depicted in the right panels. Impressively, the group delay identified and corrected the jump within 100\,ms. The lower panel offers a magnified perspective of the baseline between UT2 and UT4, with overlaid predictions from the controller's state: $\hat{\Phi}$ and $\hat{\Psi}$. Notably, at $t_2=76.9\,$s, all predictions shifted by 2200\,nm, marking the detection of the jump. To shift the OPD by one $\lambda$, the phase delay took 3 DITs (3.3\,ms), while the group delay took 150 DITs (165\,ms), attributed to the smoothing length of the observable.}
  \label{fig:GDvrPD}
\end{figure*}

\section{On-sky performances}
\label{sec:onsky}

\subsection{Hardware implementation}
\label{sec:hardware}

In November 2022, a substantial upgrade of the GRAVITY fringe-tracker hardware was carried out to increase computing power, as outlined by \citet{Abuter+2016}. As a result, the computing time was markedly reduced, opening up the possibility of implementing the updated and more complicated control algorithm. This new hardware setup is depicted in Figure~\ref{fig:Architecture}.

The upgraded real-time hardware of the controller, termed \texttt{wgvfft}, is now based on a Linux workstation. The application cycle is initiated by the arrival of detector data frames from ESO New General detector Control (NGC) workstation, facilitated through an sFPDP communication link. This application is executed on dedicated cores, with a portion of the RAM isolated from the Linux kernel, ensuring efficient access by the Reflective Memory (PCIE-5565PIORC) via DMA (Direct Memory Access).

The Kalman filter parameters ($A_L$, $Q$, $F$) are recalculated every 5 seconds by another Linux workstation (\texttt{wgvkalm}). This workstation is also fitted with a PCIE Reflective Memory card, which allows it to record the OPDs calculated by  \texttt{wgvfft} through the VLTI Reflective Memory Network (RMN) ring. This RMN ring is also used to transfer back the Kalman filter parameters to \texttt{wgvfft}.

An optional external machine, also connected to the RMN ring, can be used by another instrument to control the fringe tracker in real time. Such a workstation would have the possibility to send to \texttt{wgvfft} fringe tracking setpoints (the $U_{\rm setpoint}$). It can also be used to send the mobility flag introduced after Equation~\ref{eq:RefPD}.

The piston correction values calculated by the \texttt{wgvfft} workstation are transmitted to another workstation named \texttt{lgvttp}, using the RMN ring network. The \texttt{lgvttp} workstation operates with a Motorola CPU and is built around an mv6100 single-board computer, incorporating a VME bus for system communication. There, they are combined with external measurements of tip/tilt to form a comprehensive correction signal. This signal, in analog form, is then sent out to control four active mirrors mounted on piezoelectric tip-tilt platforms provided by Physik Instrumente.

The commissioning of the updated hardware was conducted in late 2022, as part of the GRAVITY+ upgrade \citep{2019vltt.confE..30E,2022Msngr.189...17A}, and was promptly made available to the scientific community. We could enable the white light fringe tracking a few months later, allowing the GRA4MAT mode of MATISSE \citep{Woillez+2023}. Similarly, the new Kalman filter state model, introduced in June 2023, was immediately offered to the community upon  commissioning.

\begin{figure}
  \centering
  \includegraphics[width=0.5\textwidth]{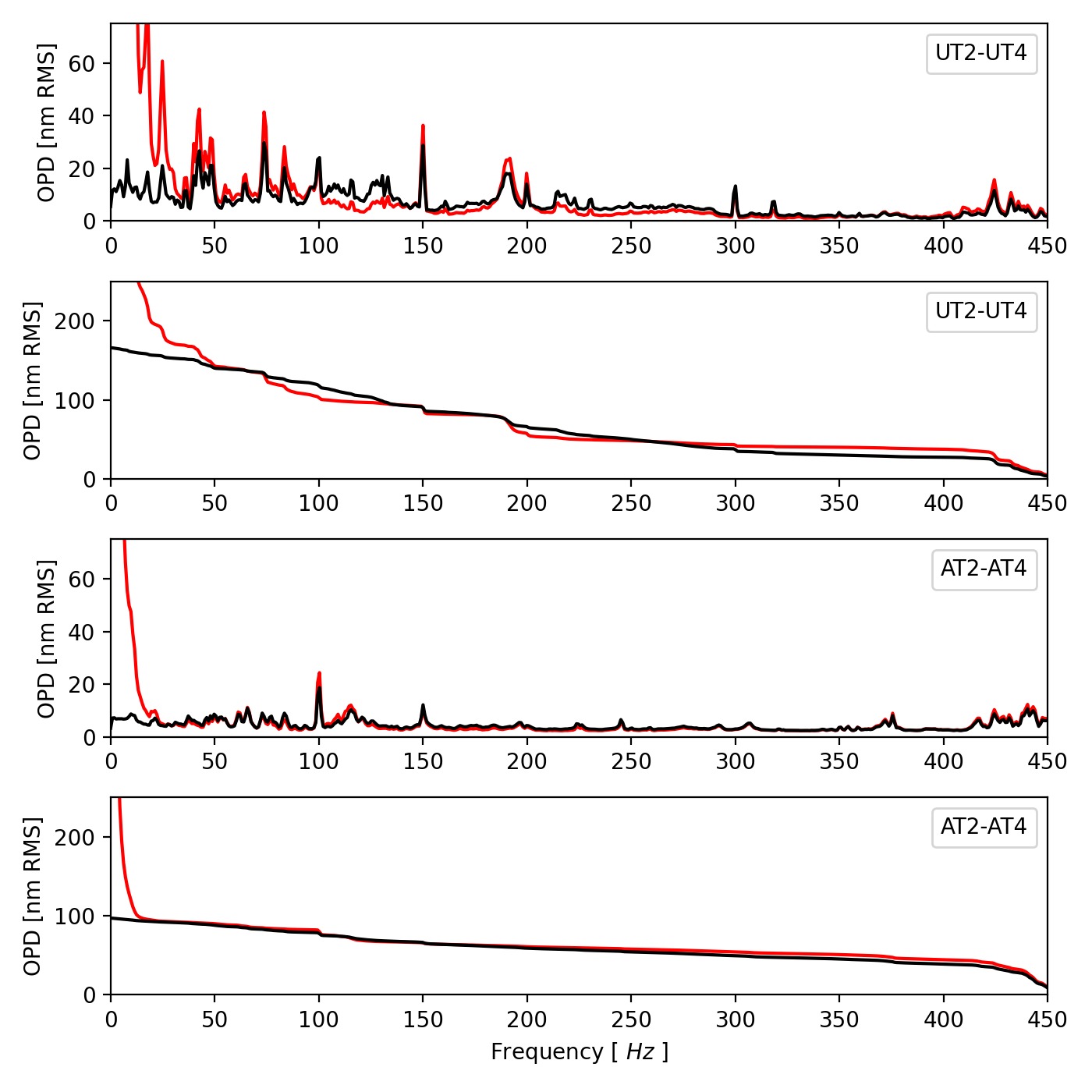}
  \caption{Spectra of the residual OPD (shown in black) and the reconstructed perturbation ($H_\phi \cdot L$, depicted in red) for two observations. The upper panels present data on HR\,8799 for one baseline from the Unit Telescopes (UTs) corresponding to the data showed in Figure~\ref{fig:GDvrPD}. 
  The standard deviation of OPD is 151~nm. The panel below showcases a dataset from the Auxilliary Telescopes (ATs) for HR\,7672 with an OPD standard deviation of 86\,nm. 
  The cumulative spectrum for these 2 dataset are also showed. The data for all baselines are presented in Appendix.~\ref{sec:psd}.}
  \label{fig:PSD}%
\end{figure}

\subsection{Performance of the group-delay control loop: analysis of a fringe jump}
\label{sec:gd}

On the 2$^{\text{nd}}$ of July 2023, we observed the exoplanet HR\,8799\,e during a scientific run of the ExoGRAVITY large programme (ESO LP 1104.C-0651). The atmospheric conditions were average, with a coherence time of 4~ms and a seeing of 0.8$^{\prime\prime}$. The planet was at a separation of 440~mas, and we observed using the "off-axis dual-field" mode of the fringe tracker. We used the roof mirror to split the field, so that all the flux from the star was injected in the FT fibre, and all the flux of the planet in the science fibre.

Due to the faint nature of the exoplanets, these observations require extended integration times on the science channel, specifically DITs of 100\,s in this example. To optimise contrast, the fringe tracker is instructed to remain on the same fringe for the entire 100\,s duration. Figure~\ref{fig:GDvrPD} shows that the FT is able to properly track the fringes on all six baselines. The residual OPD of such a constant tracking is plotted in the upper left panel of Figure~\ref{fig:GDvrPD}. HR\,8799 is a bright star of magnitude 5.24 in K-band, and is not a challenging target for the fringe tracker, which easily tracks these fringes at an interferometric SNR of $\sim{}45$. The power spectrum of the OPD residuals is shown in Figure~\ref{fig:PSD} and in Appendix.~\ref{sec:psd}.

Over a 100-second integration period, the phase-delay controller consistently tracks constant OPD values, maintaining a standard deviation of 150\,nm in the residual OPDs. Intriguingly, the group-delay residuals are notably larger, yet they manifest at lower frequencies, showcasing variations over time scales of several seconds. The peak-to-peak amplitude of these variations remains under 1000\,nm. As a result, the group-delay controller perceives no need for fringe adjustment, avoiding any false triggers for phase jump detection. Consequently, the phase-delay fringe tracker operates smoothly, successfully maintaining its tracking on the same phase.

Nonetheless, sporadic events can sometimes prompt the fringe tracker to transition abruptly from tracking one fringe to another, a full wavelength, $\lambda_0$, apart. This phenomenon, unnoticed by the phase delay, is referred to as a 'fringe jump'. A quintessential instance of this can be observed in Figure~\ref{fig:GDvrPD}. This fringe jump manifested itself at $t_1 = 76.81$ seconds across all baselines related to UT2. The lower panel of the figure gives the phase delay $\left(\Phi - \Phi_{\rm CP}\right)$, represented by solid black curves. This phase delay has already accounted for the closure phase vector, as outlined in Equation~(\ref{eq:cp}). A dashed black curve displays the estimation $\hat{\Phi}$ derived from the state parameters. Surrounding $t_1$, the OPD shows fluctuations, but the Kalman filter overlooks the phase jump, and the $\lambda_0$ deviation in the OPD is not registered by $\hat{\Phi}$. This jump manifests itself in the group delay, represented by the solid blue curve in the figure. Owing to its smoothing over 150 DITs, the group delay gradually moves to a value close to $-\lambda/2$ at $t_2 = 76.9\,$. At this point, all the UT2-related OPL state-parameters of the phase control loop undergo a shift by $\lambda_0$, induced by the group-delay controller. In the figure, this is evidenced by the abrupt $\lambda_0$ shift of both phase and group-delay state predictions ($\hat{\Phi}$ and $\hat{\Psi}$).

This example perfectly illustrates the behaviour of the group-delay controller, and shows that it required a span of 90\,ms to detect the fringe jump. This interval represents the response time of the group delay control loop, and is dictated by the group delay's smoothing length (the 150 DITs). This setting represents a balance between responsiveness and the potential for false positives due to noise\footnote{In principle, this could be a tunable parameter, which could be adjusted on a per target basis. In the case of HR~8799, given the brightness of the star and the high interferometric SNR obtained, decreasing this value would likely result in better performances. However, in the framework in which the GRAVITY FT is implemented, this is a far from easy change to make, and this value remains currently fixed at 150 DITs.}. As illustrated in the top right panels of Figure~\ref{fig:GDvrPD}, predicting this noise is challenging as it does not appear to be purely random. The value of 150 retained in the algorithm has a mostly empirical basis.

\begin{figure}
  \centering
  \includegraphics[width=0.485\textwidth]{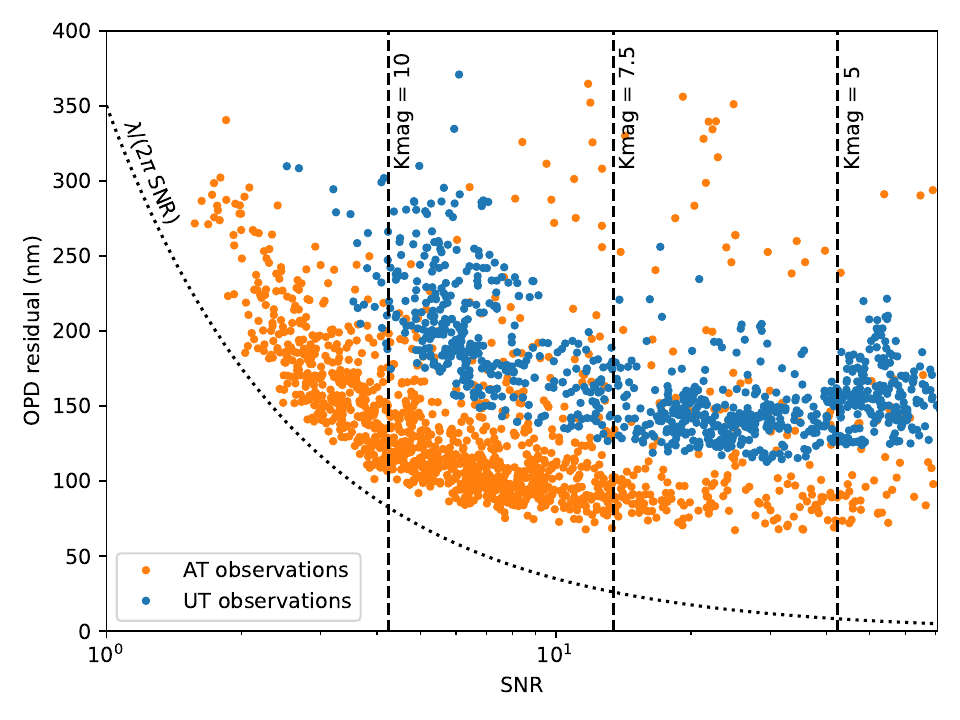}
  \includegraphics[width=0.485\textwidth]{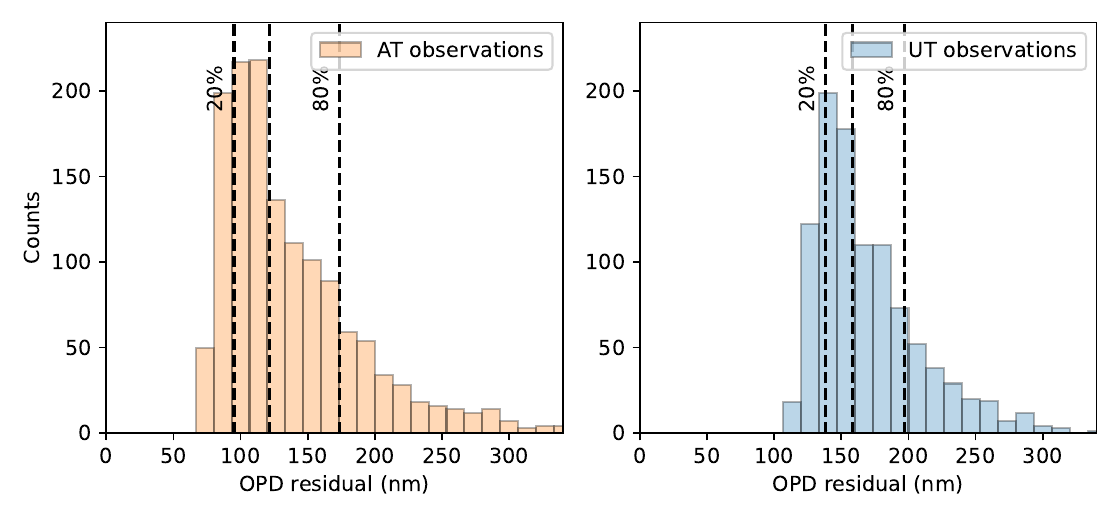}
  \caption{OPD residuals. Top panel: OPD residuals plotted against the signal-to-noise ratio per baseline for all data collected from June to August 2023. The orange dots represent ATs observations, while the blue dots denote UTs observations. The dotted curve represents the theoretical limitation imposed by the measurement noise ($\lambda/2\pi\sigma_{\rm j,k}$). The vertical dashed lines indicate the signal-to-noise ratio observed with the UTs of a star at a given star magnitude. It should be shifted by about 2.5 magnitude for the ATs. Bottom panels: histogram of the same OPD residuals as in the upper panel, for the ATs (left) and the UTs (right)}
  \label{fig:OPDresiduals}%
\end{figure}

\subsection{Performance of the phase-delay control loop: OPD Residuals}

In Paper~1, we discussed the OPD residuals observed in 2019, noting a significant dependence on the coherence time, $\tau_0$. During unfavourable conditions ($\tau_0 < 3\,$mas), values typically exceeded 380~nm. Under average conditions (3\,mas$<\tau_0 < 7\,$mas), these values ranged around 150~nm (ATs) or 250~nm (UTs), and under optimal conditions ($\tau_0 > 7\,$mas), the UTs observed values down to 220~nm. We hypothesised that the inability of the UTs to reach residuals below 220~nm, even in the best atmospheric conditions, was due to vibrations.

Over the past four years, significant efforts have been made to mitigate these vibrations. Alongside the upgraded fringe-tracker, these efforts have resulted in notable improvements. Figure~\ref{fig:PSD} display the spectrum of the OPD residuals for two targets, HR\,8799 and HR\,7672, over a single, representative, baseline. Description of the dataset and spectrum for all baselines are presented in Appendix~\ref{sec:psd}.
 The black solid curves represent the differential optical path measured by the fringe tracker ($\Phi$), while the red curves depict the differential optical path reconstructed from the actuators: $\Phi+H_\Phi\cdot{} X$.

 For the UTs, vibrations are predominant, exhibiting significant components at high frequencies, such as 20\,nm vibrations at 300\,Hz and 40\,nm at 150\,Hz. The control system strives to cancel these out, achieving partial success. However, it inadvertently reintroduces noise at other frequencies, resulting in no substantial improvement in OPD residuals down to 50~Hz. This is evident in the cumulative spectrum, where the red and black curves exhibit similar amplitudes from 50\,Hz upwards. Below 50\,Hz, the fringe tracker is highly effective, cancelling even broad-frequency vibrations around 40~Hz. For the UTs, the most detrimental range is between 50 and 100~Hz, where a forest of spectral lines challenges the controller’s ability to correct them, contributing to the observed $\approx 150$\,nm of residual. Potentially, a faster response time in the control loop could further reduce residuals within this range.

On the ATs, vibrations are not the dominant factor in the residuals, which are characterised by a white noise pattern that aligns with the theoretically expected signal-to-noise ratio. For HR\,7672, the dataset presented in Figure~\ref{fig:PSD}, the interferometric SNR is around 8, equating to a white noise level of 60\,nm. Given that residuals for different baselines range between 71 and 93\,nm, the system’s OPD residuals are close to the photon and background noise limitation. In this scenario, we observed the vibrations are minimal and do not degrade the interferometric SNR.

This behaviour of the ATs is also substantiated statistically, as demonstrated by the dataset present in the ESO archive on low signal datasets (SNR below 3). Figure~\ref{fig:OPDresiduals} shows the OPD residuals for all calibrators observed from June to August 2023. To expand the dataset, additional observations within the ExoGRAVITY large program during this period are included. The upper panel of Figure~\ref{fig:OPDresiduals} indicates that at lower fluxes for the ATs, the GRAVITY fringe tracker aligns closely with the theoretical phase error arising solely from noise, indicating a reach towards the theoretical signal-to-noise limit. However, when the signal-to-noise ratio surpasses 3 for both UTs and ATs, the performance begins to diverge from this limit, potentially due to piston noise from external factors like the atmosphere.

In conclusion, the ATs’ residuals at the 20\%, 50\%, and 80\% percentiles are 95\,nm, 120\,nm, and 170\,nm ($1\sigma$), respectively, and for the UTs, the corresponding values are 135\,nm, 150\,nm, and 185\,nm. These values are derived from the histograms in the lower panels of Figure~\ref{fig:OPDresiduals}. We observe a reduced dependence on the coherence time for performance, with optimal performance now achievable even at a low $\tau_0$ of 2\,ms.

\section{Future Prospects}
\label{sec:end}

\subsection{Areas of Improvement}

There are at least three key areas that require further attention to enhance the capabilities of fringe tracking:
\begin{itemize}
  \item Reducing OPD residuals at high SNR
  \item Enhancing signal-to-noise ratio for a given photon count
  \item Boosting sensitivity by operating at slower speeds
\end{itemize}

\textbf{Regarding OPD residuals.} The critical question is whether these can be further reduced at high signal-to-noise ratios. There appears to be potential for improvement on the UTs, as their performance, despite notable advances, still lags behind that of the ATs. Ignoring injection-related issues, the atmospheric piston impact on an 8\,m telescope should in theory be less than on a 1.8 meter telescope \citep{1995JOSAA..12.1559C}. Given that the baselines in the astrometric configuration used in our AT observations of HR~7672 are very similar to the UT baselines, the fringe tracker should have achieved better results on the UTs, and reducing OPD residuals below 80\,nm on the UTs should be feasible, with an ultimate atmospheric limitation of 5\,nm according to \citet{2022SPIE12183E..28C}. The discrepancy in performance can be possibly attributed to instrumental vibrations, particularly high above 50\,Hz (Figure~\ref{fig:PSD}), which the control loop struggle to correct, despite the fact that they seem to be properly captured by the model. Amplitude fluctuations induced by the performance of the Adaptive Optics (AO), with lower Strehl ratios observed on the UTs, are another major factor to take into account. A better control of the vibrations, as well as a better AO correction, will certainly help to further decrease the OPD residuals on the UTs.

 \textbf{Concerning signal-to-noise.} 
 Another question worthy of interest is whether the measured interferometric SNR matches the expectations given the magnitude of our targets. The observations of the bright target HR~8799 in July 2023 yielded an event rate of $n_{\rm e^-/s/UT} = 10^7$ photons per telescope per second, suggesting a combined transmission for GRAVITY and VLTI of about 1\%. The photometric SNR can be calculated as:
\begin{equation}
  \mathrm{SNR}_\mathrm{photometric} = \sqrt{\frac{4 n_{\rm e^-/s/UT} \delta t}{\gamma}}\,.
\end{equation}
\noindent{}where the factor 4 arises from the use of four telescopes, and $\gamma$ represents the efficiency of the recombination architecture, with lower values being more desirable. At minimum, $\gamma$ is inherently limited by the number of degrees of freedom in the system. Given the 16 degrees of freedom of our architecture (4 real fluxes and 6 complex coherent fluxes), an optimised beam combiner could theoretically achieve a $\gamma$ as low as 16. Assuming $\gamma = 16$, $\delta t = 0.00085\,$\footnote{The discrepancy between this integration time and the running frequency of 909~Hz is explained by the detector readout and reset time.} and given the fact that the visibility is close to one for this unresolved target, the interferometric SNR for HR~8799 should be around 46, which indeed matches our measured value of $\sim{}45$. This implies that the $\gamma$ value for GRAVITY is indeed close to 16, indicative of a well-designed combiner. Further improvements will require innovative approaches such as the Hierarchical Fringe Tracker could also potentially decrease the $\gamma$ value by reducing the system's degrees of freedom \citep{2022SPIE12183E..0IP}.

\textbf{On ultimate sensitivity.} During the June commissioning run, the exact sensitivity limit of the fringe tracker on the UTs remained undetermined, as the AO performance started to degrade at magnitudes below 10 (see UT data in Figure~\ref{fig:OPDresiduals}). The introduction of laser guide stars in GRAVITY+ is anticipated to extend the AO's limiting magnitude, potentially allowing us to ascertain the true sensitivity of the enhanced GRAVITY fringe tracker. Once the signal-to-noise ratio limit at 1~kHz is achieved, exploring a 100~Hz mode becomes a viable option for further boosting sensitivity by mitigating the detector's readout noise. In theory, as previously discussed, the theoretical atmospheric piston on an 8-meter telescope should be minimal, which could even enable operation as slow as 10~Hz. However, achieving this would necessitate a reduction in the instrumental piston caused by vibrations. Mechanical solutions, such as directly removing vibrations or compensating for them using accelerometers (as in the ongoing MANHATTAN-II project \citep{articleLieto,2018SPIE10701E..03W}), offer one approach. Alternatively, as these vibrations are intrinsic to the VLTI infrastructure, the optical path could be monitored and adjusted thanks to a dedicated metrology system, as proposed in the now-abandoned VibMET project \citep{2018SPIE10701E..03W}.

\subsection{Unexplored possibilities}

Using the GRAVITY fringe tracker alongside other instruments introduces a number of challenges in maintaining fringe tracking accuracy across different wavelengths. At a basic level, this requires the fringe tracker to avoid "fringe jumps", a critical improvement implemented by the new algorithm described in this paper. But a notable observation also pertains to the issue of varying atmospheric dispersion. The blue curves in Figure~\ref{fig:GDvrPD} display slowly changing dispersion patterns. While the GD smoothing is approximately 150\,ms, the blue curves exhibit correlations over significantly longer durations, up to 10\,s. This can be attributed to fluctuations in water vapour, as investigated in the works of \citet{2013PASP..125.1226C} and \citet{2014A&A...567A..98M}. Such signals hold relevance for infrared instruments, such as ELT/METIS \citep{2022SPIE12185E..11A}, but could also be addressed at the fringe tracker level to adjust the setpoint, ensuring stable fringes at another wavelength despite the changing dispersion.

Machine learning algorithms represent another unexplored avenue in the realm of fringe-tracking. These algorithms have the potential to integrate a broader spectrum of information compared to our current Kalman filter-based loop. Data from AO wavefront sensors and/or telescope vibration sensors, for instance, could substantially enhance fringe-tracking capabilities \citep{2022MNRAS.511.5709P}. A model, trained on a comprehensive dataset, could leverage this additional information for more accurate performance. The success observed in Adaptive Optics, particularly with reinforcement learning \citep{2022OExpr..30.2991P}, hints at the potential advancements that machine learning could bring to fringe tracking.

\section{Conclusion}
\label{sec:conclusion}

In this work, we have introduced a new approach to fringe tracking, based on a Kalman filter working in OPL state-space. This state-space is non-observable, which makes the integration of measurements trickier than when working in OPD space. In particular, this leads to a somewhat cumbersome transformation of the AR model used to predict the evolution of the OPD disturbances to OPL space. But the rewards is a reduction of dimensionality of the state-space, from 6 (the number of baselines, or OPDs) to 4 (the number of telescopes, or OPLs). In the case of the new GRAVITY FT, this allows us to propagate covariance-matrices on the state-vector, and hence to use an optimal Kalman gain, dynamically calculated at each iteration.

We described this new algorithm specifically in the context of GRAVITY, but the conceptual framework can be seemlessly extended to various other instruments. As opposed to the number of OPDs, which grows quadratically with the number of telescopes in the array, the number of OPLs only grows linearly. Therefore, the gain offered by this novel approach increases with the size of the array, thereby offering a compelling solution for future concepts such as the Planet Formation Imager \citep{Monnier2018}.

In the meantime, with this new version of the GRAVITY fringe tracker, ESO is now able to offer the interferometric community a facility fringe tracker that not only delivers competitive performance but also ensures seamless integration within the VLTI environment. This new fringe tracker is already in use with MATISSE \citep[the GRA4MAT mode, described in][]{Woillez+2023}. Beyond this, it could also accommodate visitor instruments, becoming a useful tool for a much larger community.

\begin{acknowledgements}
  The authors wish to express their gratitude to the anonymous referee for an excellent report, which contained both useful suggestions to improve the paper, corrections of important mistakes, and ideas for future work.
  GRAVITY was developed in a collaboration of the Max Planck Institute for Extraterrestrial Physics, LESIA of Paris Observatory, IPAG of Université Grenoble Alpes / CNRS, the Max Planck Institute for Astronomy, the University of Cologne, the Centro Multidisciplinar de Astrofisica from Lisbon and Porto, and the European Southern Observatory. 
  This work used observations collected at the European Southern Observatory under ESO programme 1104.C-0651 and 
  109.238N.003.
  SL, JBL, and FM acknowledge the support of the French Agence Nationale de la Recherche (ANR), under grant ANR-21-CE31-0017 (project ExoVLTI).
  DD and RL have received funding from the European Research Council (ERC) under the European Union’s Horizon 2020 research and innovation program (grant agreement CoG - 866070).
  PG acknowledges the financial support provided by FCT/Portugal through grants PTDC/FIS-AST/7002/2020 and UIDB/00099/2020. SL would like to warmly thank Sylvain Rousseau for spotting many typographical errors in the equations of Paper~1.
\end{acknowledgements}

%
   \bibliographystyle{aa} 
   \bibliography{FTbib,sylvestrebib} 


\appendix

\section{Determination of the propagation matrix $A_\base{j}{k}$ and process noise $Q_\base{j}{k}$}
\label{sec:AQ}

The goal of this appendix is to derive the $A_{j,k}$ and $Q_{j,k}$ representing the AR models in OPD space, and used in Equation~\ref{eq:A_L} and \ref{eq:Q}. Both $A_{j,k}$ and $Q_{j,k}$ matrices are obtained using the \texttt{statsmodels} Python library \citep{seabold2010statsmodels} from a sequence of 10,000 $\Phi_n$ vectors ($\approx 10\,$s at 909\,Hz). The challenge in estimating the propagation matrix lies in its need to be resilient to unwrapping errors. It should also not produce aberrant values in situations where the fringes are undetected or lost.

To address this, the matrices are determined based on phase differences rather than the phases themselves. Explicitly, for each baseline, we compute $\Delta \phi_{j,k,n} = \left[ \phi_{j,k,n} - \phi_{j,k,n-1}\right]_{\text{mod }\lambda_0}$. To ensure reliable default operation during low signal-to-noise ratios on this baseline (e.g., when fringes are absent or lost), we set $\Delta \phi_{j,k,n} = 0$.

The statsmodel's time series analysis is then employed to fit an AR model to the 9,999 $\Delta \phi_{j,k,n}$ values, one baseline at a time. We use the \texttt{statsmodels.tsa.ar\_model.AutoReg} class to fit an AR model of order 22 using conditional maximum likelihood. The AR(22) model was chosen as it strikes a balance between computational efficiency and its ability to fit low frequencies. With 22 samples at 909~Hz, the minimum frequency and resolution of the predictor can be expected to be around 40~Hz. In practice, though, we find that the resolution is much better, and we are able to fit for frequencies as low as $\sim{}10\,\mathrm{Hz}$, in line with the fact that AR models are notoriously better in terms of resolution than Fourier-Transform based algorithm \citep{Quirk1983, Tary2014}.

The \texttt{statsmodels.tsa.ar\_model.AutoReg} class fit gives six vectors of 22 values for the AR model, $g_{l,j,k}$, along with six scale parameters $q_{l,j,k}$. The AR(22) model of the phase difference is then integrated to give an AR(23) model $g'$ of the phase itself, using:
\begin{align}
  &g^\prime_{0,j,k} = g_{0,j,k} , \\
  &g^\prime_{23,j,k} = -g_{22,j,k} ,\\
  &g^\prime_{l,j,k} = g_{l,j,k}-g_{l-1,j,k} \quad \text{for } l \text{ between } 1 \text{ and } 22.
\end{align}
This ensures the stationnarity of our model since $\sum_{l=0}^{23} g^\prime_{l,j,k} = 0$. The propagation matrices representing the AR model are then:
 \begin{equation}
  A_{j,k} =
  \begin{bmatrix}
  g^\prime_{0,j,k} & g^\prime_{1,j,k} & \cdots & g^\prime_{21,j,k} & g^\prime_{22,j,k} & 0 & \cdots & 0 \\
  1 & 0 & \cdots & 0 & 0 & 0 & \cdots  & 0 \\
  0 & 1 & 0 & \cdots & 0 & 0 & \cdots & 0 \\
  0 & 0 & 1 & \cdots & 0 & 0 & \cdots & 0 \\
  \vdots & \vdots & \vdots & \ddots & \vdots & \vdots & \vdots & \vdots \\
  0 & 0 & 0 & \cdots & 1 & 0 & \cdots & 0 \\
  0 & 0 & 0 & \cdots & 0 & 1 & 0 & 0 \\
  0 & 0 & 0 & \cdots & 0 & 0 & 1 & 0 \\
  \end{bmatrix},
  \end{equation}

These matrices are of dimension $150\times150$, but only the first 23 values and the sub-diagonal are non-zero. The AR(23) could be represented by a $23\times{}23$ matrix, but the extra-dimensions are required by our state-space model, which also needs to account for the group-delay. 

The matrix $Q_{j,k}$ is also of dimension $150\times150$. It is more straightforwardly generated from the scale parameter $q_{0,jk}$ which correspond to the variance of the residuals in the \verb|statsmodels| fit:
\begin{equation}
   Q_{j,k} = \begin{bmatrix}
    q_{0,j,k} &  0  & \dots & 0 \\
    0 &  0  & \dots &0 \\
    \vdots &  \vdots  & \ddots & 0 \\
    0 &  \dots  & \dots & 0 
   \end{bmatrix} 
 \end{equation}

\section{Power spectrum density}
\label{sec:psd}

\begin{figure*}
  \centering
  \includegraphics[width=\textwidth]{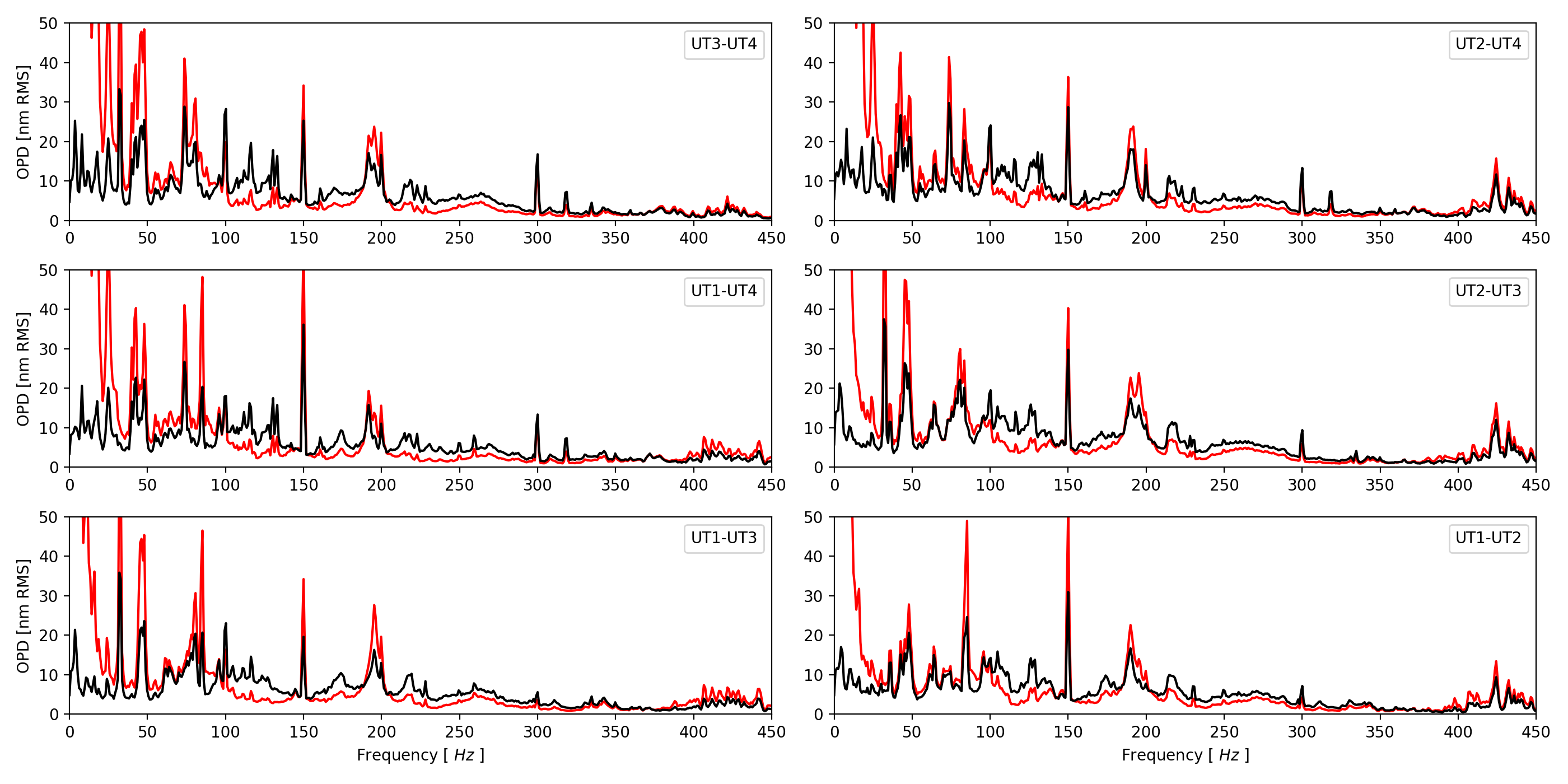}
  \includegraphics[width=\textwidth]{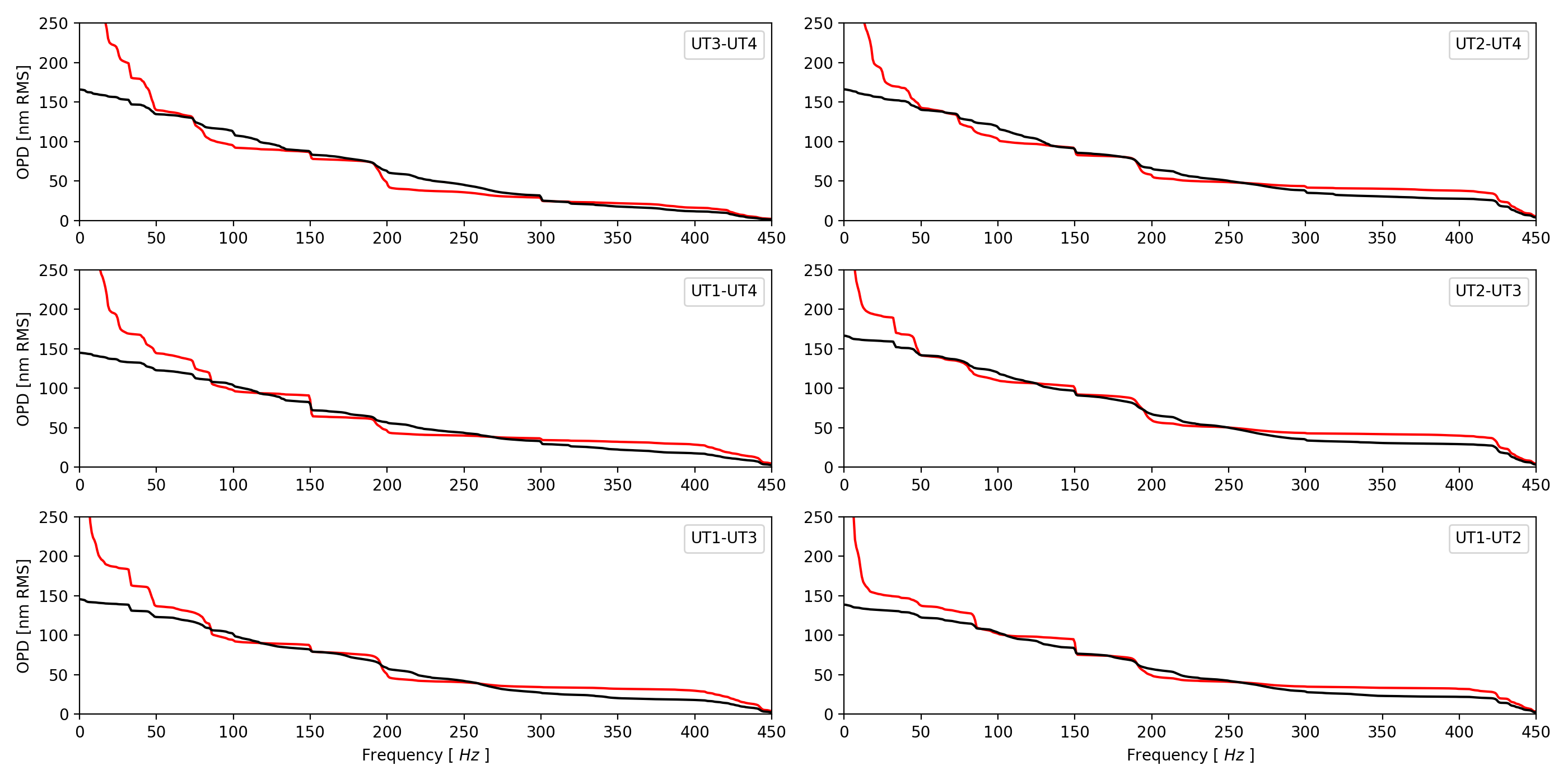}
  \caption{Power spectrum density (PSD) and cumulative sum of the PSD of phase residuals ($\vec \Phi_{n}$) for target HR~8799. The red curves represent the pseudo open-loop values, signifying the scenario without a fringe tracker ($H_\phi \cdot L$). The six lower plots show the reverse cumulative sum of the power spectrum. Over 100\,s of a single scientific DIT, the residual standard deviation is 158\,nm (UT3-UT4),
   150\,nm (UT2-UT4), 133\,nm (UT1-UT4), 152\,nm (UT2-UT3), 138\,nm (UT1-UT3) and 124\,nm (UT1-UT2). }
  \label{fig:PSD_UT}%
  \end{figure*}

\begin{figure*}
    \centering
    \includegraphics[width=\textwidth]{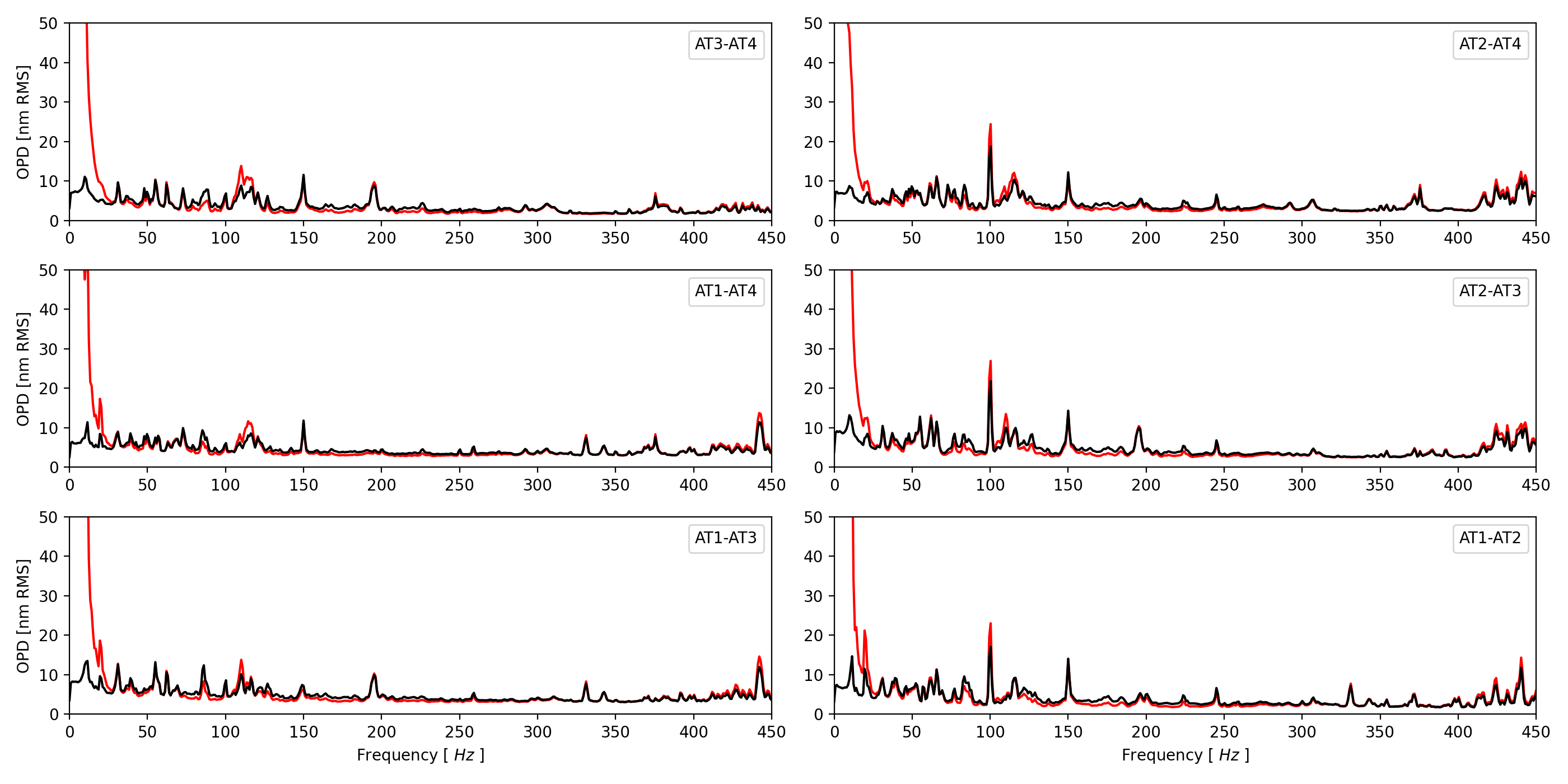}
    \includegraphics[width=\textwidth]{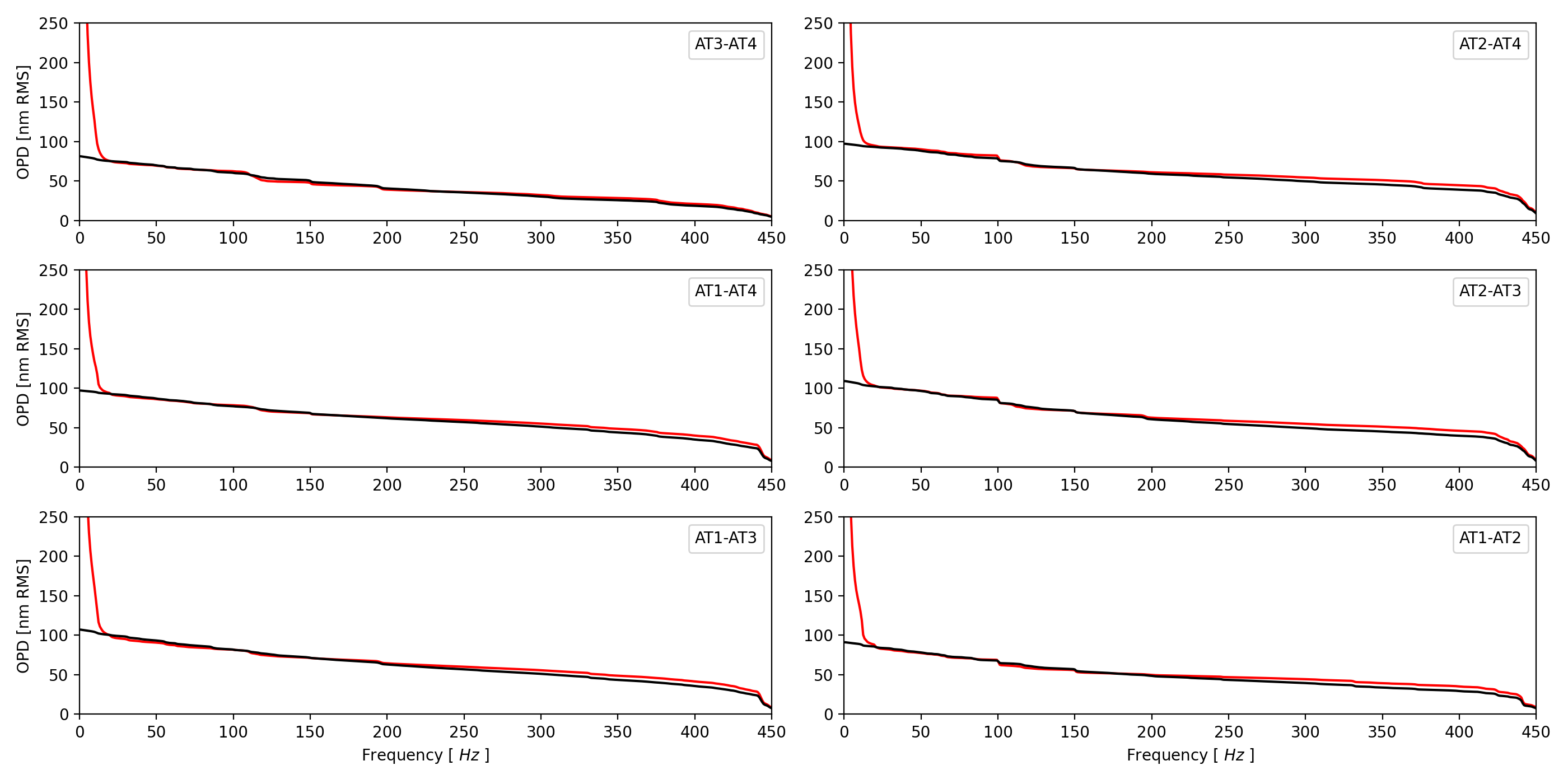}
    \caption{Power spectrum density (PSD) and cumulative sum of the PSD of phase residuals ($\vec \Phi_{n}$) for target HR~7672. The red curves represent the pseudo open-loop values, signifying the scenario without a fringe tracker ($H_\phi \cdot L$). The six lower plots show the reverse cumulative sum of the power spectrum. Over 100\,s of a single scientific DIT, the residual standard deviation is 71\,nm (UT3-UT4),
    86\,nm (AT2-AT4), 86\,nm (AT1-AT4), 93\,nm (AT2-AT3), 93\,nm (AT1-AT3) and 80\,nm (AT1-AT2).}
    \label{fig:PSD_AT}%
\end{figure*}

In Figures~\ref{fig:PSD_UT} and~\ref{fig:PSD_AT}, we present the Fourier transforms of 100 seconds of fringe tracking OPD, $\Phi$, acquired with the UTs and the ATs, respectively. The first dataset was obtained using the UTs to observe HR\,8799 (Kmag$=5.44$) on July 2, with an effective K magnitude of 4.4. The second dataset was captured using the ATs, in the "astrometric" configuration (A0-G1-J2-K0) to observe HR\,7672 (Kmag$=4.4$) on July 21. The exact file IDs for these observations are ``GRAVI.2023-07-02T09:52:22.630'' and ``GRAVI.2023-07-21T06:16:19.030''. Atmospheric conditions were good during the HR\,8799 observations with a seeing of 0.8'' and wind speed of 3.9 to 4.6~m/s, and excellent during the HR\,7672 observations with a seeing of 0.4'' and wind speed of $\sim{}2.8~\mathrm{m/s}$. We used Welch's method from the scipy library to compute the spectrum and cumulative spectrum of the OPD. The overall standard deviations of the OPDs are provided in the figure captions.

\end{document}